\documentclass[12pt,twoside]{article}
\usepackage[mathscr]{eucal}
\usepackage{amsmath,amsfonts,amssymb,amsthm}
\usepackage{hyperref}
\usepackage[matrix,arrow]{xy}
\usepackage{times}
\usepackage{graphicx}
\usepackage{multibox}
\usepackage{dcolumn}

\def\d_Vphi{\text{d}_V\hspace{-0.06em}\phi}
\def\d_Vphibar{\text{d}_V\hspace{-0.06em}\bar\phi}
\def\d_Vxi{\text{d}_V\hspace{-0.06em}\xi}

\def\be{\begin{eqnarray}}
\def\ee{\end{eqnarray}}
\def\beann{\begin{eqnarray*}}
\def\eeann{\end{eqnarray*}}
\def\beq{\begin{equation}}
\def\eeq{\end{equation}}
\def\ba{\begin{array}}
\def\ea{\end{array}}
\def\ben{\begin{enumerate}}
\def\een{\end{enumerate}}
\def\bea{\begin{eqnarray}}
\def\eea{\end{eqnarray}}

\def\5{\bar }
\def\6{\partial }
\def\7{\hat }
\def\4{\tilde }
\advance\textheight\topskip
\renewcommand{\tilde}{\widetilde}
\renewcommand{\hat}{\widehat}

\newcommand{\bref}[1]{\textbf{\ref{#1}}}

\newcommand{\dd}{\partial}
\renewcommand{\d}{\partial}

\newcommand{\binner}[2]{%
  {\langle}\kern-4.15pt{\langle}#1{,}\,#2{\rangle}\kern-4.15pt{\rangle}}

\newcommand{\half}{\mathchoice{%
    \ffrac{1}{2}}{\frac{1}{2}}{\frac{1}{2}}{\frac{1}{2}}}

\newcommand{\ffrac}[2]{\raisebox{.5pt}%
  {\footnotesize$\displaystyle\frac{#1}{#2}$}\kern1pt}

\newcommand{\ddl}[2]{\ffrac{\dd #1}{\dd #2}}

\newcommand{\vddl}[2]{{\ffrac{\delta #1}{\delta #2}}}

\newcommand{\bundle}{\boldsymbol}

\def\cA{\mathcal{A}}
\def\cB{\mathcal{B}}

\def\cE{\mathcal{E}}
\def\cF{\mathcal{F}}
\def\cG{\mathcal{G}}
\def\cH{\mathcal{H}}

\def\cL{\mathcal{L}}
\def\cM{\mathcal{M}}

\def\cP{\mathcal{P}}
\def\cQ{\mathcal{Q}}

\newcommand{\lc}{\boldsymbol{\varepsilon}}
\numberwithin{equation}{section} \makeatletter
\@addtoreset{equation}{section}

\begin{document}

\def\mytitle{Dyons with potentials: duality and black hole thermodynamics}

\pagestyle{myheadings}
\markboth{\textsc{\small Barnich, Gomberoff}}{%
  \textsc{\small Dyons with potentials}}
\addtolength{\headsep}{4pt}

\begin{flushright}\small
ULB-TH/07-14
\end{flushright}

\begin{centering}

  \vspace{1cm}

  \textbf{\Large{\mytitle}}

  \vspace{1.5cm}

  {\large Glenn Barnich$^{a}$}

\vspace{.5cm}

\begin{minipage}{.9\textwidth}\small \it \begin{center}
   Physique Th\'eorique et Math\'ematique, Universit\'e Libre de
   Bruxelles\\ and \\ International Solvay Institutes, \\ Campus
   Plaine C.P. 231, B-1050 Bruxelles, Belgium \end{center}
\end{minipage}

\vspace{.5cm}

{\large Andr\'es Gomberoff}

\vspace{.5cm}

\begin{minipage}{.9\textwidth}\small \it \begin{center}
   Universidad Nacional Andr\'es Bello, Av. Rep\'ublica 239, Santiago, Chile
\end{center}
\end{minipage}

\end{centering}

\vspace{1cm}

\begin{center}
  \begin{minipage}{.9\textwidth}
    \textsc{Abstract}. A modified version of the double potential
    formalism for the electrodynamics of dyons is constructed. Besides
    the two vector potentials, this manifestly duality invariant
    formulation involves four additional potentials, scalar potentials
    which appear as Lagrange multipliers for the electric and magnetic
    Gauss constraints and potentials for the longitudinal electric and
    magnetic fields. In this framework, a static dyon appears as a
    Coulomb-like solution without string singularities. Dirac strings
    are needed only for the Lorentz force law, not for Maxwell's
    equations. The magnetic charge no longer appears as a topological
    conservation law but as a surface integral on a par with electric
    charge. The theory is generalized to curved space. As in flat
    space, the string singularities of dyonic black holes are
    resolved. As a consequence all singularities are protected by the
    horizon and the thermodynamics is shown to follow from standard
    arguments in the grand canonical ensemble.
  \end{minipage}
\end{center}

\vfill

\noindent
\mbox{}
\raisebox{-3\baselineskip}{
\parbox{\textwidth}{\mbox{}\hrulefill\\[-4pt]}}
{\scriptsize$^a$Senior Research Associate of the Fund for
  Scientific Research-FNRS.}

\thispagestyle{empty}
\newpage

\begin{small}
{\addtolength{\parskip}{-1.5pt}
 \tableofcontents}
\end{small}
\newpage

\section{Introduction}
\label{sec:introduction}

Reissner-Nordstr{\o}m black holes with both electric and magnetic
charge
\begin{eqnarray}
  \label{eq:20}
   ds^2&=&- N^2 dt^2+
  N^{-2}dr^2+
  r^2(d\theta^2+\sin^2\theta d\phi^2),\nonumber\\
  && N=\sqrt{1-\frac{2M}{r}+\frac{Q^2+P^2}{r^2}},\\
 A&=&-\frac{Q}{r}dt +P(1-\cos\theta)d\phi,\label{eq:20a}
\end{eqnarray}
are generally excluded in a discussion of uniqueness theorems and
geometric derivations of the first law because the gauge potential is
singular along a string that intersects the horizon and goes to
infinity \cite{Carter:1972,Sudarsky:1992ty}. Exceptions can be found
in \cite{Heusler:1996} for stationary and axisymmetric perturbations
and in \cite{Copsey:2005se} where dipole charge contributions to the
first law for five dimensional black ring solutions are investigated
by dealing directly with divergent potentials on the horizon. Dyonic
solutions were also excluded in an investigation of duality of
electric and magnetic black holes using Euclidean methods
\cite{Hawking:1995ap}.

Nevertheless, for variations of the three parameters the first law
\begin{eqnarray}
  \label{eq:34}
  \delta M=\frac{\kappa}{8\pi}\delta\cA+\phi_H\delta Q+\psi_H\delta
  P,
\end{eqnarray}
where
\begin{eqnarray}
&& \Delta=M^2-(Q^2+P^2),\quad r_\pm=M\pm\sqrt{\Delta}  \\
&& \kappa=\frac{r_+-r_-}{2r_+^2},\quad \cA=8\pi\big[M^2-\frac{Q^2+P^2}{2}+
M\sqrt{\Delta}\big],\label{eq:35a}\\
&& \phi_H=\frac{Q}{r_+},\quad \psi_H=\frac{P}{r_+},\label{eq:35}
\end{eqnarray}
can easily be inferred from the purely electric case by using a
duality argument. Furthermore electric-magnetic black hole duality has
been extended to the case of dyons in the canonical ensemble by using
the manifestly duality invariant double potential formalism
\cite{Deser:1997xu}. In its original version \cite{Deser:1976iy}, this
formalism involves as dynamical degrees of freedom two vector
potentials. An independent rederivation \cite{Schwarz:1993vs} has been
written with two additional scalar potentials which are spurious
because they appear only as a part of a total derivative of the
action. In the black hole context \cite{Deser:1997xu}, coupling to
external static sources can be made either through fixed strings,
spherically symmetric nondynamical longitudinal fields, or
intermediate combinations. Finally, the coupling to dynamical dyons
with the help of dynamical strings has been studied in detail in
\cite{Deser:1997mz,Deser:1997se}, including a proof of equivalence
with Dirac's original theory \cite{Dirac:1931kp,Dirac:1948um} and a
derivation of the appropriate quantization condition
\cite{PhysRev.173.1536,PhysRev.176.1489}.

What we will do in this paper is introduce potentials for longitudinal
components of electric and magnetic fields. This has the effect of
making the two scalar potentials non-spurious as they now appear as
the Lagrange multipliers for the divergence constraints on electric
and magnetic fields. We thus increase the redundancy of the
description in such a way as to have twice as much gauge invariance as
in standard Maxwell theory.

Now, taking into account all the results described above, this
extension of the double potential formalism is rather straightforward
and seems hardly worth the effort. We beg to differ.

First of all, the electric and magnetic potentials produced by a
static dyon both appear as Coulomb-like solutions in a single,
manifestly duality invariant formulation without any stringlike
singularities. In this framework Dirac strings are only needed in
order to produce the correct Lorentz force law from an action
principle for dynamical point-particle dyons.

In curved space, the new formulation is ideally suited for the
description of black hole dyons. As in flat space, their string
singularityis resolved and a geometric derivation of the first law can
be done along standard lines because all singularities are now
protected by the horizon. This is a direct consequence of the
intriguing transmutation into a surface integral of the magnetic
charge which appears as a topological conservation law in the standard
approach. Since there is no quantization condition on magnetic or on
electric charge for a single dyon and because of the presence in the
formalism of both chemical potentials, thermodynamics and Euclidean
computations can be performed in the grand canonical ensemble, thus
circumventing arguments of \cite{Sudarsky:1992ty,Hawking:1995ap}.

In the next section, we discuss our formulation in Minkowski space in
the case of fixed external
sources. Section~\bref{sec:dynam-point-part} is devoted to Dirac
strings and dynamical point-particle dyons. We finally write down and
analyze the appropriate action for curved space and discuss
applications in the context of black hole physics.

\section{Extended double potential formalism in flat space}
\label{sec:extend-doube-potent}

In this section we present an action principle for electromagnetism in
the presence of electric and magnetic sources which is manifestly
duality invariant. Both electric and magnetic Gauss constraints are
dynamical and appear in the action with their corresponding Lagrange
multipliers. For a static dyon, the solution of the field equations is
Coulomb-like, both in the electric and the magnetic sector. We show
that the theory can be gauge fixed so as to coincide with standard
electromagnetism and conclude the section by showing that Lorentz
invariance, while not manifest, is nevertheless realized through
canonical generators very much as in the standard Hamiltonian
formulation of electromagnetism. This suggests, as we will explicitly
show in the last section, that the theory can be generalized to curved
space.

\subsection{Action, duality, and gauge symmetries}
\label{sec:action-symmetries}

The dynamical fields of the theory are $A^a_\mu,C^a$, $a=1,2$. Here,
$A^a_\mu\equiv(A_\mu,Z_\mu)$ are the standard and new potentials. The
additional fields $C^a\equiv(C,Y)$ make up the longitudinal parts of
magnetic and electric fields $\vec B^a\equiv (\vec B,\vec E)$
according to
 \begin{eqnarray}
 \vec B^a=\vec \nabla \times \vec A^a +\vec \nabla C^a.
\label{eq:36}
\end{eqnarray}
The external magnetic and electric currents $j^{a\mu}\equiv
(k^\mu,j^\mu)$ are conserved, $\partial_\mu j^{a\mu}=0$. In this
section, we assume that they correspond to the currents produced by a
single point-particle dyon. We consider the action
\begin{equation}\label{action0}
I[A_\mu^a,C^a]= I_M[A_\mu^a,C^a]+I_I[A_\mu^a;j^{a\mu}],
\end{equation}
where
\begin{equation}
  \label{eq:37}
  I_M[A_\mu^a,C^a]=\half\int d^4x\,
  \Big[\epsilon_{ab}(\vec B^{a}+\vec\nabla C^a)\cdot
(\d_0 \vec A^b- \vec \nabla A^b_0) -\vec B^{a}\cdot \vec B_a\Big],
\end{equation}
is the substitute for the usual Maxwell action and
\begin{eqnarray}
  \label{eq:23}
I_I[A_\mu^a;j^{a\mu}]=\int d^4x\, \epsilon_{ab} A^a_\mu j^{b \mu}
\end{eqnarray}
is the ``interaction'' action. Here $\epsilon_{ab}$ is skew-symmetric
with $\epsilon_{12}=1$, and indices $a,b,\dots$ raised and lowered
with the Kronecker delta.  The action (\ref{action0}) is manifestly
invariant under simultaneous duality rotations on $(A^a_\mu,
C^a;j^{a\mu})$
\begin{equation}
\delta_D A^a_\mu =  \epsilon^{ab}A_{b\mu} ,
\ \ \ \  \delta_D C^a = \epsilon^{ab}C_b ,
\ \ \ \  \delta_D j^{a\mu} = \epsilon^{ab}j_b^\mu .\label{duality}
\end{equation}
It is also gauge invariant under
\begin{eqnarray}
\delta_\lambda A^a_\mu=\partial_\mu\lambda^a,\qquad
\delta_\lambda C^a=0\label{eq:45}.
\end{eqnarray}

\subsection{Equations of motion and point-particle dyon}
\label{sec:equat-moti-point}

The Euler-Lagrange equations of motion associated with \eqref{eq:37}
are easily shown to be equivalent to Maxwell's equation with magnetic
and electric currents. Indeed, variations with respect to $A^a_0$ give
the constraints
\begin{eqnarray}
\vec \nabla  \cdot\vec B^a\equiv\nabla^2 C^a=j^{0a}.\label{eq:33}
\end{eqnarray}
Variations with respect to $C^a$ imply the equations
\begin{eqnarray}
  \label{eq:39}
  \nabla^2 C_a=\epsilon_{ab}(\vec\nabla \cdot\d_0{\vec A}^b-\nabla^2 A^b_0).
\end{eqnarray}
The fields $A^0_a,C^a$ are auxiliary in the sense that, under suitably
boundary conditions at spatial infinity, their equations of motion can
be solved for $A^0_a,C^a$ in terms of all other fields, without the
need for intial conditions.

Variations with respect to $\vec A^a$ yield Maxwell's equations in the form
\begin{eqnarray}
  \label{eq:40}
 -\epsilon_{ab}\d_0 {\vec B}^b+\vec\nabla \times \vec B_a=\epsilon_{ab}
 \vec j^{b}.
\end{eqnarray}

As a consequence, if the electromagnetic field tensor $F$ is expressed
in the usual way in terms of electric and magnetic fields,
$F_{0i}=-B^2_i$, $F_{ij}=\epsilon_{ijk}B^{1k}$, it follows
that both $dF$ and $d{}^*F$ vanish outside of sources on account of the
Euler-Lagrange equations of motion.

In the case of a single point-particle dyon at the origin with charges
$Q^a\equiv(P,Q)$, for example,
\begin{eqnarray}
j^{a\mu}(x)=4\pi Q^a\delta^\mu_0\delta^3(x),
\label{eq:46}
\end{eqnarray}
instead of \eqref{eq:20a}, Maxwell's equations in the above form are
now solved by
\begin{eqnarray}
  \label{eq:47}
  A^a=-\frac{\epsilon^{ab}Q_b}{r}dt, \quad
C^a=-\frac{Q^a}{r}.
\end{eqnarray}
This solution resolves the string-singularity of the standard
formulation. It is unique in the transverse gauge $\vec \nabla \cdot
\vec A^a =0 $ with vanishing boundary conditions on $A^a_\mu, C^a$.

\subsection{Canonical structure and degrees of freedom}
\label{sec:poiss-brack-degr}

By using integrations by parts and decomposing $\vec A^a=\vec
A^{aT}+\vec \nabla M^a$, with $M^a=(M_A,M_Z)$, the free action
\eqref{eq:37} can be written in the form
\begin{multline}
  \label{eq:37a}
  I_M[\vec A^{aT},A_0^a,M^a,C^a] =\int d^4x\,
\Big[-\vec \nabla\times \vec Z_T \cdot
\d_0 \vec A_T+\nabla^2 Y \d_0 M_A-\nabla^2 C \d_0 M_Z
  -\\
  -\half \vec E\cdot\vec E-\half\vec B\cdot\vec B -A_0\nabla^2 Y+
  Z_0\nabla^2 C\Big],
\end{multline}
where $\vec E=\vec \nabla\times\vec Z+\vec \nabla Y$, $\vec B=\vec
\nabla\times\vec A+\vec \nabla C$.  This shows that the canonically
conjugate pairs are $(\vec A_T,-\vec\nabla\times \vec Z_T)$,
$(\nabla^2 Y, M_A)$ and $( -\nabla^2 C, M_Z)$ so that there are $4$
conjugate pairs per spacetime point.

Variation with respect to the Lagrange multipliers $Z_0$ imposes the
first class constraint $\nabla^2 C=0$. Partial gauge fixing to the
standard covariant description can be achieved by requiring the
longitudinal part of the second vector potentials to vanish,
$M_Z=0$, and gives back the usual Hamiltonian
description of electromagnetism. Complete gauge fixation is then
achieved, as usual, by solving the electric Gauss constraint $\nabla^2
Y=0$ associated with the Lagrange multiplier $A_0$ together with the
gauge condition $M_A=0$. The gauge fixed theory
contains $2$ physical degrees of freedom per spacetime point described
by the transverse vector potential $\vec A^T$ and its canonically
conjugate variable $-\vec E^T=-\vec\nabla\times\vec Z^T$, as it
should.

For later use, we note that
\begin{eqnarray}
  \label{eq:88}
  \{A^{ai}(x),B^{bj}(x^\prime)\}=-\epsilon^{ab}\delta^{ij}
\delta^3(x,x^\prime), \quad  \{C^{a}(x),B^{bj}(x^\prime)\}=0,\nonumber\\
\{M^a(x),
C^b(x^\prime)\}=\epsilon^{ab}\nabla^{-2}\delta^3(x,x^\prime),\quad
\{B^{ai}(x),B^{bj}(x^\prime)\}=\epsilon^{ab}\epsilon^{ijk}
\d_k\delta^3(x,x^\prime).
\end{eqnarray}

\subsection{Duality, gauge, and Poincar\'e generators}
\label{sec:poinc-gener-their}

The Hamiltonian and constraints associated with the first order action
$I_M[A^a_\mu,C^a]$ are
\begin{eqnarray}
  \label{eq:89}
  H=\int d^3x\, \half \vec B^a\cdot\vec B^a,\quad g_a=
  \epsilon_{ab}\vec \nabla \cdot\vec B^b,
\end{eqnarray}
The duality generator is the $SO(2)$ Chern-Simons term
\cite{Deser:1976iy} suitably extended to the longitudinal potentials,
\begin{eqnarray}
  \label{eq:90}
  D=-\half \int d^3x\,(\vec B^a+\vec \nabla C^a)\cdot\vec A_a.
\end{eqnarray}
It commutes with the Hamiltonian and the other Poincar\'e generators
introduced below, but is only weakly gauge invariant,
\begin{eqnarray}
  \label{eq:91}
  \{g_a,D\}=\epsilon_{ab}g^b.
\end{eqnarray}
The duality transformations \eqref{duality} on the canonical variables
$A_i^a,C^a$ are generated through $\delta_D A_i^a=\{A_i^a,D\}$,
$\delta_D C^a=\{C^a,D\}$. The extension to the Lagrange multipliers is
dictated by \eqref{eq:90} and the requirement that the first order
action $I_M[A^a_\mu,C^a]$ is invariant. In the same way, the gauge
transformations $\delta_\lambda$ in \eqref{eq:45} are generated by
\begin{eqnarray}
  \label{eq:94}
  \Upsilon[\lambda]=\int d^3x\, g_a\lambda^a.
\end{eqnarray}
In this expression, the generators are smeared with the arbitrary
functions $\lambda^a$ defining the gauge transformation in
\eqref{eq:45}.

A general  Poincar\'e generator may be written as
\begin{eqnarray}
T(\omega,a)=\half
\omega_{\mu\nu}J^{\mu\nu}-a_\mu P^\mu,\label{eq:16}.
\end{eqnarray}
Here $J^{\mu\nu}$ and $P^\mu$ are the individual Poincar\'e
generators, and $\omega_{\mu\nu}$, $a_\mu$ the corresponding
parameters defining the transformation. The generator of time
translations is the Hamiltonian, $P^0=H$. The Lorentz generators may
be decomposed as
\begin{eqnarray}
  \label{eq:74}
  \half
\omega_{\mu\nu}J^{\mu\nu}=\half\omega_{ij}\epsilon^{ijk}J_k-\omega_{i0}K^i.
\end{eqnarray}
The Poincar\'e generators are related to the symmetric energy-momentum
tensor with complete electric and magnetic fields as follows:
\begin{eqnarray}
 T^{00}=\half (\vec E^2+\vec B^2),\quad T^{i
  0}=(\vec E\times \vec B)^i,\\
P^\mu=\int d^3x\, T^{\mu 0},\quad  J^{\prime \mu\nu}=-\int d^3x\, (x^\mu T^{\nu
  0}-x^\nu T^{\mu 0}), \label{eq:97}
\end{eqnarray}
where $J^\prime_k=J_k$ and $\vec K^\prime= \vec K-x^0 \vec P$,
\begin{eqnarray}
  \label{eq:92}
  \vec P=-\half \int d^3x\, \epsilon_{ab}\vec B^a\times\vec B^b,&&
  \vec J=\int d^3x\, \epsilon_{ab}\vec B^a(\vec x\cdot \vec
  B^b),\nonumber\\
  &&\hspace*{-3cm} \vec K=
  \int d^3x\,\vec x \,(\half \vec B^a\cdot\vec B^a).
\end{eqnarray}
One can then show by direct computation of the Poisson brackets that
these generators form a representation of the Poincar\'e algebra, up
to terms involving the constraints. We will prove this explicitly in
Sec.~\bref{sec:append-poiss-algebra} and show that
\begin{eqnarray}
  \label{eq:xx}
  \{T(\omega,a),T(\theta,b)\} = T
  ([\omega,\theta],\omega b-\theta a)
  + \Upsilon[[\xi,\eta]_B],
\end{eqnarray}
where
\begin{eqnarray}
  \xi(\omega,a)^\mu &=& -\omega^\mu_{\phantom{\mu}i}x^i+a^\mu,\cr
 \eta(\theta,b)^\mu &=& -\theta^\mu_{\phantom{\mu}i}x^i+b^\mu,\cr
[\xi,\eta]^a_B &=& B^{ai}\epsilon_{ijk}\xi^j\eta^k-
\epsilon^{ac}B_{ci}
  (\xi^0\eta^i -\eta^0\xi^i)\label{eq:102}.
\end{eqnarray}
If we then define
\begin{eqnarray}
a^\prime_i=a_i+\omega_{0i}x^0\label{eq:105}
\end{eqnarray}
and all other parameters unchanged, the conserved Noether charges
generating the Poincar\'e transformations as canonical transformations
on the fields are
\begin{eqnarray}
  \label{eq:104}
  Q(\omega,a)=T(\omega,a^\prime).
\end{eqnarray}
Indeed, deriving \eqref{eq:xx} in terms of $a^\prime(a)$ with
respect to $b_0$ and putting $\theta=0$ gives
\begin{eqnarray}
  \label{eq:103}
  \{H,Q(\omega,a)\}=\ddl{}{t}Q(\omega,a)-\int d^3x\, g_a\epsilon^{ab}
  B_{bi}\xi^i(\omega,a^\prime).
\end{eqnarray}
As a consequence, the Noether charges are conserved on the constraint
surface, as they should, and the Poincar\'e transformations of the
canonical variables, $\delta_Q C^a=\{C^a,Q\}$ and $\delta_Q
A^a_i=\{A^a_i,Q\}$, can be extended to the Lagrange multipliers so as
to leave the action invariant. Explicitly, with the understanding that
$\xi=\xi(\omega,a^\prime(a))$,
\begin{eqnarray}
  \label{eq:101}
  \delta_Q C^a&=&0,\ \\
\delta_Q A^a_i&=&\d_i\lambda_Q^a
-\epsilon^{ab} B_{bi}\xi^0-\epsilon_{ijk}\xi^jB^{ak},\\
\delta_Q B^a_i&=&-\epsilon^{ijk}\d_j(
  \epsilon^{ab}B_{bk}\xi^0)-\d_j(B^{aj}\xi^i)+
\d_j(B^{ai}\xi^j),
\\ \delta_Q A_0^a&=&\d_0 \lambda_Q^a+ \epsilon^{ab}B_{bi}\xi^i,
\end{eqnarray}
where
\begin{eqnarray}
  \lambda_Q^a=-\epsilon^{ab}\nabla^{-2}\d_i(B^i_b\xi^0)+
\nabla^{-2}\d_i(\epsilon^{ijk} B^a_j \xi_k).
\end{eqnarray}

\section{Dynamical point-particle dyons}
\label{sec:dynam-point-part}

We show in this section that for sources that correspond to dynamical
point-particle dyons, a consistent action principle that makes the
dyons evolve according to the Lorentz force law needs Dirac-type
strings and requires a veto giving rise to the standard quantization
condition. We then show equivalence with Dirac's original, manifestly
Lorentz invariant formulation.

\subsection{Dyons and Dirac strings}
\label{sec:dyons-dirac-strings}

We begin by reviewing the use of Dirac strings in the theory of
magnetic monopoles.  Let us first fix conventions. Define
$\epsilon_{a_1\dots a_n}=\epsilon^{a_1\dots a_n}$ to be totally
skew-symmetric with $\epsilon_{1\dots n}=1$. The Levi-Civita tensor is
$\lc_{a_1\dots a_n}=\sqrt{|g|} \epsilon_{a_1\dots a_n}$. Indices on
this tensor are raised with the metric, which implies that
$\lc^{a_1\dots a_n}=\frac{(-)^\sigma}{\sqrt{|g|}} \epsilon^{a_1\dots
  a_n}$ where $\sigma$ is the signature of the metric. Our convention
for the dual is $({}^*\!\omega^p)_{a_1\dots a_{n-1}}=\frac{1}{p!}
\omega^{b_1\dots b_p}\lc_{b_1\dots b_p a_1\dots a_{n-p}}$.

Consider a (d+1)-dimensional surface $\Sigma_{d+1}$ in
flat 4-dimensional spacetime parameterized by $(\tau, \sigma_1,
\ldots, \sigma_d)$,
$$
x^{\mu} = v^\mu(\tau, \sigma_1, \ldots, \sigma_d).
$$
Associated with this surface, define the $d+1$ form $H_{\Sigma_{d+1}}$
with contravariant components
\begin{equation}
  H_{\Sigma}^{\mu_1 \cdots \mu_{d+1}}(x) = \int_\Sigma \
\delta^{(4)} (x-v) dv^{\mu_1} \wedge \cdots \wedge dv^{\mu_{d+1}}.
 \end{equation}
It is straightforward to show that if $\partial \Sigma$ is the
 boundary of $\Sigma$, then,
\begin{equation}
\label{dh}
d {}^*\! H_{\Sigma_{d+1}} = {}^*\!H_{\partial \Sigma_{d+1}}.
\end{equation}
In the Dirac theory, the worldline $\Gamma: x^\mu=z^\mu(\tau)$ of a
magnetic pole of charge $g$ defines the magnetic current
\begin{equation}
j_{mag}^\mu= g H_\Gamma^\mu.
\end{equation}
The worldline is the boundary of the worldsheet of a Dirac string
$\Sigma: x^\mu=y^\mu(\tau,\sigma)$. Hence, if
$G^{\mu\nu}=gH_{\Sigma}^{\mu\nu}$,
\begin{equation}
d^*\!G= {}^*\!j_{mag}.\label{eq:34f}
\end{equation}
Dirac defines the electromagnetic field by
\begin{equation} \label{diracF}
F = da + {}^*\!G
\end{equation}
and gets the desired modified Bianchi identity $dF={}^*\!j_{mag}$.
Note that we have used the lowercase $a_\mu$ for the electromagnetic
potential here.  This is to distinguish it from the potentials $A^a_i$
in our formalism ( see Eq. \eqref{eq:36}).  In particular, $A^1_i$ in
our formulation is not equal to $a_i$, which arises in other
two-potential formulations to be discussed below.

In the Dirac formulation, the theory has an extra gauge symmetry
associated with the freedom of arbitrarily choosing the position of the
strings while keeping its boundary (worldline of monopole) fixed. To
see this, consider the displacement of a string defined by
\begin{equation}
\label{disp}
x^{\mu} = w^\mu(\tau, \sigma, \lambda).
\end{equation}
where the initial string worldsheet $\Sigma$ is at $\lambda=0$ and the
final, $\Sigma^\prime$, at $\lambda=1$. The boundary of the 3-dimensional
surface $\Upsilon$ defined by \eqref{disp} is $\Delta \Sigma = \Sigma
-\Sigma^\prime$. Hence, if
\begin{equation}
\Delta H^{\mu\nu}_{\Sigma} = H^{\mu\nu}_{\Sigma}- H^{\mu \nu}_{\Sigma'},
\end{equation}
then from \eqref{dh},
\begin{equation}\label{DJs}
 {}^*\! \Delta G = d{}^*\!K,
\end{equation}
where $K^{\alpha\beta\gamma}=gH_{\Upsilon}^{\alpha\beta\gamma}$.
Therefore, we see that the electromagnetic field $F$ in \eqref{diracF}
is invariant under the displacement of the string if, while moving the
string, we also vary $a$ by
\begin{equation}\label{da}
    \Delta a = - {}^*\!K.
\end{equation}
The Dirac action, which depends on the string only through
$F^{\mu\nu}$ is invariant under this gauge symmetry, up to the anomaly
that gives rise to the quantization condition, which will be explained
in more detail below when discussing the double potential formalism.

In a manifestly duality invariant theory, magnetic and electric
charges are treated on the same footing. In general one considers $n$
dynamical dyons with magnetic and electric charges
$q^a_n\equiv(g_n,e_n)$. The current is then defined
as
\begin{equation}
j^{a\mu}(x) = \sum_n q_n^a H_{\Gamma_n}^\mu(x)=\sum_{n} q^{a}_n \int_{\Gamma_n}
  \delta^4(x-z_n)dz_n^\mu,\label{eq:current}
\end{equation}
where the sum in $n$ is over the worldlines $\Gamma_n$ of every dyon
of charge $q^a_n$ [parameterized by $z_n^\mu(\tau)$ with an arbitrary
parameter $\tau$]. For the Dirac strings attached to them, we define
\begin{equation}
\label{sc}
G^{a\mu\nu}(x) = \sum_n q_n^a
H_{\Sigma_n}^{\mu\nu}(x)=\sum_{n}q_n^a\int_{\Sigma_n}\delta^4(x-y_n)dy^\mu_n\wedge
dy^\nu_n,
\end{equation}
where $\Sigma_n$ is the worldsheet of the Dirac string whose boundary
is $\Gamma_n$ [parameterized by  $y_n^\mu(\tau,\sigma$)
 with arbitrary parameters $\tau$ and $\sigma$]. 
The analogs of Eqs.~\eqref{eq:34f} and
\eqref{DJs} in this case are
\begin{equation}
\label{DJ}
d{}^*\! G^a = {}^*\! j^a,\quad {}^*\! \Delta G^a = d{}^*\!K^a,
\end{equation}
where
\begin{equation}
K^{a \  \alpha\beta\gamma} = \sum_n q^a_n H^{\alpha\beta\gamma}_{\Upsilon_n} \ ,
\end{equation}
and $\Upsilon_n$ is the surface defined by the displacement of the
string attached to the dyon $q^a_n$.

When splitting space and time, as in the different manifestly duality
invariant formulations, it is convenient to also split the space and
time components of the string currents, defining
\begin{equation}\label{ab}
  \alpha^a_i = \frac{1}{2} \epsilon_{ijk}\ G^{a \ jk} =
  {}^*\!G^a_{0i} \ ,
\ \ \ \ \ \ \ \ \
  \beta^{ai} =  G^{a \ 0i}=\frac{1}{2}\epsilon^{ijk} \ {}^*\!G^a_{jk}
\end{equation}
Explicitly,
\begin{eqnarray}
  \label{eq:6}
  \vec \alpha^a&=& \sum_n q^a_n
  \int_{\Sigma_n} \delta^4(x-y_n)   \half d\vec y_n\times \wedge d\vec y_n,\\
  \vec \beta^a&=& \sum_n q_n^a
  \int_{\Sigma_n} \delta^4(x-y_n)    dy^0_n \wedge d\vec y_n,  \label{eq:6b}
\end{eqnarray}
where $(d\vec y_n\times \wedge d\vec y_n)_i =
\epsilon_{ijk}dy^i \wedge dy^j$ and the first identity in
\eqref{DJ} becomes
\begin{eqnarray}
  \vec \nabla \cdot\vec \beta^a=j^{a 0},\ \
  \vec \nabla\times\vec \alpha^a -\d_0 \vec
  \beta^a = \vec j^a. \label{eq:21a}
\end{eqnarray}
It is also convenient to work with the dual of
$K^{a\ \alpha\beta\gamma}$, the one-form $v_\alpha^a$. In terms of it,
we may derive the way the vectors $\vec \alpha^a$ and $\vec \beta^a$
in \eqref{ab} transform under displacement of the strings. Using the
second identity in \eqref{DJ},
\begin{eqnarray}
\Delta \alpha^a_i &=& {}^*\!\Delta G^a_{0i} = (dv^a)_{0i} =
\partial_0 v^a_i - \partial_i v_0^a  , \label{dalpha}\\
\Delta \beta^{a i} &=& \frac{1}{2}\epsilon^{ijk}
\ {}^*\!\Delta G^a_{jk} = \frac{1}{2}\epsilon^{ijk}
\ (dv^a)_{jk} = (\vec \nabla \times \vec v^a)^i . \label{dbeta}
\end{eqnarray}

For dynamical dyons, the action must be supplemented with the kinetic
term
\begin{eqnarray}
I_k[z^\mu_n]=-\sum_{n}\int_{\Gamma_n}\sqrt{-dz_n^\mu dz_{n\mu}}.\label{eq:48}
\end{eqnarray}
The total action $I'$ that includes the dynamics of the dyons and
produces the correct Lorentz force law is
\begin{multline}\label{action1}
    I'[A^a_\mu,C^a,y^\mu_n] = I_M +I_I+ I_k+\\+ \half\int d^4 x\, \epsilon_{ab}
    \left[ 2\vec\nabla C^a \vec \alpha^b - \vec\beta^a\vec\alpha^b
-  \vec \beta^a\nabla^{-2}\vec\nabla\times\d_0\vec \beta^b \right].
\end{multline}
The constraints (\ref{eq:33}) and the electromagnetic
Eq.~(\ref{eq:40}) are clearly unchanged, for the extra piece in the
action does not depend on $A_\mu^a$. The equations obtained from the
variation of $C^a$ are modified with respect to the result of \eqref{eq:39}
to
\begin{eqnarray}
  \label{eq:9}
  \nabla^2 C_a=\epsilon_{ab}\left(\vec\nabla \cdot[\d_0{\vec
    A}^b+\vec \alpha^b]-\nabla^2 A^b_0\right).
\end{eqnarray}
For later use we note that, applying $\vec\nabla\times$ to
(\ref{eq:40}) and using (\ref{eq:9}), together with the boundary
condition that $\vec B^a$ falls off at least as fast as $r^{-1}$ at
infinity,
\begin{eqnarray}
  \label{eq:18}
  \vec B_a&\approx&\epsilon_{ab} \Big(\d_0\vec
  A^b-\vec\nabla A_0^b+\vec\alpha^b+\nabla^{-2}\vec\nabla\times
  \d_0\vec \beta^b\Big).
\end{eqnarray}

As a side remark, we also note that the definitions
\begin{eqnarray}
{\bf F}^{a}_{\mu\nu}&=&\d_\mu A^a_\nu-\d_\nu A_\mu^a+{}^*\cG^{a}_{\mu\nu},\quad
{}^*\cG^{a}_{ij}=\epsilon_{ijk}\d^k C^a, \\
{}^*\cG^{a}_{0 i}&=&\alpha^a_i+
\nabla^{-2} (\vec\nabla\times \d_0\vec
 \beta^a)_i. \label{eq:24a}
\end{eqnarray}
are such that
\begin{eqnarray}
  \label{eq:87}
  B_{ai}=\half\epsilon_{ijk}{\bf F}^{jk}_a=
{}^*{\bf F}_{a0i}.
\end{eqnarray}
Furthermore, they allow us to write the on-shell Eqs.~\eqref{eq:18} as
\begin{eqnarray}
 \label{eq:26}
 B_{ai}\approx
 \epsilon_{ab}{\bf F}^{b}_{0i}=-\half\epsilon_{ijk}\epsilon_{ab}{}^*{\bf F}^{jkb},
\end{eqnarray}
while the equations of motion for $A^a_\mu$ take the covariant form
\begin{eqnarray}
\d_\nu {\bf F}^{\mu\nu}_a=\epsilon_{ab}j^{b\mu},\quad
\d_\nu{}^*{\bf F}^{\mu\nu}_a=-j_a^\mu.
\label{eq:26a}
\end{eqnarray}

In the case of a single dyon, one assumes without loss of generality
that the string terms in the last line of \eqref{action1} are
absent. Indeed, in this case one can perform a duality rotation so
that, say, the magnetic charge vanishes. The string terms then reduce
to $-\int d^4x \vec\nabla C^1 \alpha^2$ and can be dropped because
they only affect the irrelevant auxiliary equation used to determine
$A_0^2$. This justifies {\em a posteriori} the coupling to the sources
considered in the first section.

\subsection{Lorentz force law and veto}
\label{sec:lorentz-force-law}

We still need to vary $y^\mu,z^\mu$ in action \eqref{action1} in order
to derive the Lorentz force law.  We will see below that in order to
obtain it, we need to impose the so called ``Dirac veto.'' This demand
was introduced by Dirac in his original treatment of magnetic
monopoles \cite{Dirac:1948um} to obtain the desired classical
equations. It consists of the requirement that no electric charge can
touch a Dirac string. At the quantum level, Dirac showed that the veto
modifies the topology of phase space, giving rise to his celebrated
quantization condition.  In our formalism the Dirac veto is required
as well, as we show below. The difference resides in that here we will
need to ask that no dyon can touch the string of any other dyon. This
generalized version of the Dirac veto was also used in
\cite{Deser:1997mz}.

Variations of $I_I$ with respect to $z^\mu_n$ give
\begin{multline}
  \label{eq:5}
  \delta_z
  I_I=\sum_n\epsilon_{ab}q^b_n\int_{\Gamma_n}\Big((
\d_0 \vec A^a-\vec\nabla A_0^a)
\cdot(\delta z^0_nd\vec z_n-\delta\vec z_ndz^0_n)
+\\
+ (\vec \nabla \times\vec A^a)\cdot (\delta\vec z_n\times d\vec
z_n)\Big).
\end{multline}

Before varying $I_M$ with respect to $y^\mu_n$, we establish
the following identities. For all smooth vector fields $\vec V^a$,
$\vec W^a$ one has
\begin{multline}
  \label{eq:7}
  \int d^4x\, \vec V_b \delta_y\vec\alpha^b=
\sum_nq^b_n\Bigg[\\\int_{\Gamma_n}\vec V_b\cdot (\delta \vec
z_n\times  d\vec z_n)+\int_{\Sigma_n}\Bigg(\vec\nabla\cdot\vec V_b\,
\delta \vec y_n\cdot\half
(d\vec y_n\times\wedge d\vec y_n)-\\-
\d_0\vec V_b\cdot\Big(\delta \vec y_n\times (dy^0_n\wedge
d\vec y_n)-\delta y^0_n\half (d\vec y_n\times\wedge d\vec y_n)\Big)
\Bigg) \Bigg]
\end{multline}
and
\begin{multline}
  \label{eq:7a}
  \int d^4x\, \delta_y\vec\beta^a\vec W_a=
\sum_nq^a_n\Bigg[\int_{\Gamma_n}(\delta z^0_nd\vec z_n-\delta\vec
z_n dz^0_n)\cdot \vec W_a+\\
+\int_{\Sigma_n}\Big(\delta \vec y_n\times (dy^0_n\wedge
d\vec y_n)-\delta y^0_n\half (d\vec y_n\times\wedge d\vec y_n)
\cdot (\vec\nabla \times\vec W_a)\Big) \Bigg].
\end{multline}
The variation of $I_M$ with respect to $y^\mu_n$ may be computed by
specializing for the fields
\begin{eqnarray}
  \label{eq:8}
  \vec V_b&=&\half\epsilon_{ab}( 2 \vec \nabla C^a-\vec
  \beta^a), \label{aaa} \\
  \vec W_a&=& -\half\epsilon_{ab}(\vec\alpha^b+
2\nabla^{-2} \vec\nabla\times\d_0\vec \beta^b ). \label{bbb}
\end{eqnarray}
Combining all terms,
\begin{multline}
  \label{eq:42}
  \delta _z I_I+\delta_y I_M
=\sum_nq^a_n\Bigg[\int_{\Gamma_n}\Bigg(\Big(\vec W_a-
\epsilon_{ab}(
\d_0 \vec A^b-\vec\nabla A_0^b)\Big)
\cdot(\delta z^0_nd\vec z_n-\delta\vec z_ndz^0_n)
+\\+
\Big(\vec V_a-\epsilon_{ab} (\vec \nabla \times\vec A^b)\Big)
\cdot (\delta\vec z_n\times d\vec
z_n)\Bigg) +\int_{\Sigma_n}\Bigg(
\vec\nabla\cdot\vec V_a\,
\delta \vec y_n\cdot\half
(d\vec y_n\times\wedge d\vec y_n)+\\+
 (\vec\nabla \times\vec W_a-\d_0 \vec V_a)\cdot
\Big(\delta \vec y_n\times (dy^0_n\wedge
d\vec y_n)-\delta y^0_n\half (d\vec y_n\times\wedge d\vec y_n)
\Big)\Bigg)
 \Bigg].
\end{multline}
Now, taking the divergence of \eqref{aaa} and making use of the first
identity in \eqref{eq:21a} and the constraints \eqref{eq:33} one gets,
\begin{eqnarray}
  \label{eq:14}
  \vec\nabla \cdot \vec V_a=\half\epsilon_{ab}j^{b0}.
\end{eqnarray}
It follows that the second term in \eqref{eq:42} vanishes provided the
string attached to dyon $n$ does not cross any other dyon (Dirac
veto). This is due to the fact that the Dirac veto ensures that
$j^{\mu a}=0$ on the worldsheet of the strings. Similarly, from
\eqref{aaa}, \eqref{bbb} and the identities \eqref{eq:21a} it is
straightforward to show that
\begin{eqnarray}
  \label{eq:10}
\vec\nabla\times\vec W_a-\d_0 \vec V_a=-\half\epsilon_{ab} \vec
j^b.
\end{eqnarray}
Hence, the last term in \eqref{eq:42} also vanishes on account of
Dirac's veto. The string piece in the first and second terms of
\eqref{eq:42} again vanish because of the veto. This may be seen from
the fact that \eqref{aaa} may be written as,
\begin{eqnarray}
  \label{eq:11}
\vec V_a-\epsilon_{ab} (\vec \nabla \times\vec
A^b)=-\epsilon_{ab}\vec B^b+\half\epsilon_{ab}
\beta^b,
\end{eqnarray}
and therefore, due to the veto, the integral on the worldline of a
dyon only sees the first term. In the same way, using \eqref{eq:18},
\begin{eqnarray}
  \label{eq:12}
  \vec W_a-\epsilon_{ab}( \d_0 \vec A^b-\vec\nabla A_0^b)
=-\vec B_a+\half\epsilon_{ab}\vec\alpha^b,
\end{eqnarray}
and the second term vanishes on the worldline of a dyon.

Combining the remaining terms with those from the variation of $I_k$,
extremization of the total action now implies the Lorentz force law
\begin{eqnarray}
\label{eq:41d}
 m_n\frac{d}{d\tau}\left(
\frac{\frac{d z^0_{n}}{d\tau}}{\sqrt{-\frac{dz_{n\mu}}{d\tau}
\frac{dz^\mu_n}{d\tau}}}\right)&=& q^a_n\vec B_a(z_n)\cdot\frac{d\vec
    z_n}{d\tau},\\
m_n\frac{d}{d\tau}\left(\frac{\frac{d \vec z_{n}}{d\tau}}
{\sqrt{-\frac{dz_{n\mu}}{d\tau}
\frac{dz^\mu_n}{d\tau}}}\right)&=& q^a_n\vec B_a(z_n)\cdot\frac{d
    z^0_n}{d\tau}+\frac{d\vec z_n}{ds}\times\vec
B^a(z_n)\epsilon_{ab}q^b_n,\label{eq:41a}
\end{eqnarray}
as it should.

\subsection{Equivalence with Dirac's covariant formulation and
  quantization condition}
\label{sec:equiv-with-diracs}

We end this section by showing that the theory presented above is
equivalent to Dirac's theory. This shows that the theory with dyons is
Lorentz invariant. We will actually show that our action
(\ref{action1}) is equivalent to an action found in \cite{Deser:1997mz}
which, in turn, has been shown in \cite{Deser:1997se} to be equivalent
to a generalization of Dirac's covariant formulation allowing for
dyons.

Explicitly, this action reads
\begin{equation}\label{d}
  \bar I[\vec{a}^a,y^\mu_n] =
  \half \int d^4x\,\left[\epsilon_{ab} \vec b^a (\d_0 \vec  a^b+\vec
    \alpha^b)-\vec b^a\cdot\vec b_a
    + \epsilon_{ab}\vec a^a \cdot\vec   j^b \right]+ I_k,
\end{equation}
where
\begin{equation} \label{b}
\vec{b}^a = \vec \nabla \times \vec a^a + \vec \beta^a.
\end{equation}
This formulation makes use of Dirac strings in the same way as our
formulation does. That is, each dyon $q_n$ is attached to a string
parameterized by $y^\mu_n(\tau,\sigma)$. The quantities $\vec\alpha^a$
and $\vec\beta^a$ appearing in \eqref{d} are the same ones as defined
in Eqs. \eqref{eq:6}, \eqref{eq:6b} above and satisfy the identities
\eqref{eq:21a}. Note that on account of these identities, the
longitudinal part of $\vec a$ drops out of this action principle.  The
field $\vec b^a$ is the magnetic/electric field appearing in Maxwell's
equations. It must, therefore, be the same as our $\vec B^a$.

In this formulation, Gauss' law appears as an identity on taking
the divergence of $\vec b^a$ in (\ref{b}). The field $\vec \nabla \times
\vec a^a$ is transversal but has a stringlike singularity which is
removed by $\vec\beta^a$. In our formulation $\vec B^a$ has two
nonsingular pieces, namely, the transverse and longitudinal
components of it. To show equivalence, we decompose $\vec
\beta^a$ accordingly so that
\begin{equation} \label{bb} \vec{b}^a = \vec \nabla \times \vec a^a +
  \vec \beta^a = \vec \nabla \times (\vec a^a -\nabla^{-2} \vec \nabla
  \times \vec \beta^a ) + \vec \nabla \nabla^{-2} \vec \nabla \cdot
  \vec \beta^a .
\end{equation}
{}From the constraints (\ref{eq:33}) and the first identity in
(\ref{eq:21a}), the longitudinal piece is precisely $\vec \nabla
C^a$.  We are then lead to the following identifications:
\begin{eqnarray}\label{Aa}
  \vec A^a &=& \vec a^a -\nabla^{-2} \vec \nabla \times \vec \beta^a, \\
  C^a &=& \nabla^{-2} \vec \nabla \cdot
  \vec \beta^a  \label{Aa2}.
\end{eqnarray}
(The first equation is true up to an irrelevant longitudinal field, in
order for $\vec B^a=\vec b^a$ to hold.)

To establish equivalence is now straightforward. We start with the
action (\ref{action1}) of our formulation.  Assuming vanishing
boundary conditions at spatial infinity, $(A^a_0,C^a)$ are auxiliary
fields because their equations of motions (\ref{eq:33}) and
\eqref{eq:9} can be used to algebraically determine them in terms of
the other fields. We can thus solve for them in action
(\ref{action1}). Then we use (\ref{Aa}) to write $\vec A^a$ in terms
of $\vec a^a$, and after a bit of algebra involving the identities
(\ref{eq:21a}) we get precisely action (\ref{d}).

In the double potential formulation of \cite{Deser:1997mz}, i.e., for
action \eqref{d}, the symmetry corresponding to shifts in the
string is realized by transforming $\vec \alpha^a$ and $\vec \beta^a$
in \eqref{dalpha}, \eqref{dbeta}, with $v_0^a=0$ and
\begin{equation}\label{Da}
\Delta \vec a^a = -\vec v^a .
\end{equation}
Let us do that for the case in which there are only two dyons, $q^a$,
$\bar q^a$.  We will only change the position of the string attached
to $q^a$.  Varying action \eqref{d} and using the identity $\vec
\nabla \cdot \vec b^a = j^{a0}$ we get
\begin{equation} \label{daction}
\delta{\bar I} = \frac{1}{2}\epsilon_{ab} \int d^4 x j^{a\mu} v^{b}_\mu.
\end{equation}
This is zero unless the worldline of dyon ${\bar q}^a$ crosses the
3-dimensional manifold $\Upsilon$ swept by the string attached to
$q^a$. In that case the variation is
\begin{equation}\label{dII}
\delta{\bar I} = \frac{1}{2}\epsilon_{ab} q^a {\bar q}^b .
\end{equation}
This will not affect the quantum mechanical system if the variation is
proportional to $2\pi\hbar n$, for some integer $n$. This leads us to
the Dirac-Schwinger-Zwanziger quantization condition
\begin{eqnarray}
  \label{eq:32}
  \bar eg-e\bar g=2\pi n\hbar,
\end{eqnarray}
up to a factor of $1/2$. This factor is removed by a careful analysis
of the topology of the system. We will not discuss this here. More
details can be found in \cite{Deser:1997mz, Deser:1997se}.

Finally, we study what happens in our formulation.  First, let us
compute how the field $\vec A^a$ transforms under the movement of the
string. From Eq.  \eqref{Aa}, \eqref{dbeta} and \eqref{Da} we get
\eqref{Aa},
\begin{equation}
\label{dA}
 \Delta \vec A^a  = - \vec v^a -
\nabla^{-2} \vec \nabla \times \nabla \times \vec v^a = -
\vec \nabla \left( \nabla^{-2} \vec\nabla \cdot \vec v^a \right).
\end{equation}
If we now take
\begin{equation}\label{da0}
  \Delta A^a_0 = -v_0^a= -\d_0\left( \nabla^{-2} \vec\nabla \cdot \vec v^a
  \right)+\d_0\left( \nabla^{-2} \vec\nabla \cdot \vec v^a
  \right)-v_0^a,
\end{equation}
the variation defined by \eqref{dA} and the first term of \eqref{da0}
is a gauge transformation of the form \eqref{eq:45} which leaves
action $I^\prime$ in (\ref{action1}) invariant.  We thus only need to
compute the variation under the movement of the strings and the second
part of \eqref{da0}. Using identities (\ref{eq:21a}) one obtains
precisely the same result as in the previous case, namely, the
right-hand side of Eq.~\eqref{daction}. The argument leading to the
quantization condition is therefore the same.

\section{Extended double potential formalism in curved space}
\label{sec:black-hole-therm}

We generalize the first order action to curved spacetimes and discuss
the canonical and gauge structure of the theory, including
diffeomorphism invariance. In particular, we show that the standard
algebra of surface deformations of the purely gravitational case now
involves both Gauss-type constraints with structure functions
depending on electric and magnetic fields. We proceed to the equations
of motion deriving from the generalized action principle and show that
they are equivalent to the covariant Einstein-Maxwell equations. We
then show how the string singularity of the Reissner-Nordstr{\o}m
dyonic black hole solution gets resolved in our formalism. We compute
the electric and magnetic surface integrals following the
Regge-Teitelboim approach, discuss how they appear in a geometric
derivation of the first law and in the Euclidean approach to black
hole thermodynamics. Finally, we apply these results to the resolved
Reissner-Nordstr{\o}m black hole.

\subsection{Action and canonical structure}
\label{sec:acti-canon-struct}

The first order action $I_M$ can be generalized to curved spacetimes.
We consider a globally hyperbolic spacetime, foliated by a spacelike
family of hypersurfaces, each labeled by the value of a timelike
coordinate $t$. The induced metric on each surface is $g_{ij}(t)$. We
follow the conventions of MTW \cite{Misner:1970aa}, chapter 21, where
spatial indices are lowered and raised with the 3-metric $g_{ij}$ and
$g$ is its determinant.  We denote by $\epsilon_{ijk}$ the completely
antisymmetric symbol, which differs from the $[ijk]$ notation used in
MTW.

Adapting the results derived in
\cite{Arnowitt:1962aa,Deser:1976iy,Schwarz:1993vs,Deser:1997xu} and
defining $\cB^{ai}=\epsilon^{ijk}\partial_jA^a_k+\sqrt g\partial^i
C^a$, we get the following manifestly duality invariant action in the
absence of sources:
\begin{multline}
  \label{eq:17}
  I_M[A^a_\mu,C^a,g_{ij},N,N^i]=\frac{1}{8\pi}\int
  d^4x\,\Big[(\cB^{ai}+\sqrt g\d^i C^a)
\epsilon_{ab}(\d_0 A^b_i-\d_i A_0^b)
-\\-\frac{N}{\sqrt g}\cB^{i}_a\cB_i^a-\epsilon_{ab}
\epsilon_{ijk}N^i\cB^{aj}\cB^{bk}\Big],
\end{multline}
where $N=(-{}^{(4)}\!g^{00})^{1/2}$ and $N^i= {}^{(4)}\!g^{ij}\
{}^{(4)}\!g_{0i}$ are the lapse and shift functions and ${}^{(4)}\!
g_{\mu\nu}$ is the 4-dimensional metric.

We are interested in solutions to the equations of motion derived from
$I=I_{ADM}+I_M$, where $ I_{ADM}$ is the first order action for pure
general relativity.  Introducing the collective notation
$z^A=(g_{ij},\pi^{ij},A^a_i,C^a)$ for the different fields in our system,
this action principle takes the form
\begin{eqnarray}
  \label{eq:25}
I[z,u]=\int d^4x\, [ a_A(z)\d_0 z^A -u^\alpha\gamma_\alpha],\\
 a_A(z)\d_0z^A= \frac{\pi^{ij}}{16\pi}\d_0 g_{ij}-\frac{\cE^i}{4\pi}\d_0
A_i + \frac{\sqrt g\d^i C}{4\pi}\d_0 Z_i.
\end{eqnarray}
The constraints $\gamma_\alpha\equiv(\cH_\perp,\cH_i, \cG_a)$
are associated with the Lagrange multipliers $u^\alpha\equiv
(N,N^i,A^a_0)$ and given by\footnote{Note the misprint in eq. (21.116)
  of \cite{Misner:1970aa}, where there should be no lapse function on
  the right-hand side.}
\begin{eqnarray}
\cH_\perp=\frac{1}{16\pi}(\cH^{ADM}_\perp+{\cH^{mat}_\perp}), \quad
\cH_i=\frac{1}{16\pi}(\cH^{ADM}_i+\cH^{mat}_i),\quad
 \cG_a=\frac{1}{4\pi}\epsilon_{ab}\d_i\cB^{bi}\label{eq:55},
\end{eqnarray}
where $\cH^{ADM}_\perp,\cH^{ADM}_i$ are given in
\cite{Arnowitt:1962aa,Misner:1970aa} and
\begin{eqnarray}
  \label{eq:17a}
\cH^{mat}_\perp=\frac{2g_{ij}}{\sqrt{g}}\cB^i_a\cB^{aj},\quad
  \cH^{mat}_i= 2\epsilon_{ab}\epsilon_{ijk}\cB^{aj}\cB^{bk}.
\end{eqnarray}

The first two sets of constraints in \eqref{eq:55} above are the
gravitational Hamiltonian and momentum constraints, while the last set
are the two electromagnetic Gauss constraints.

In order to disentangle the canonical structure we begin by writing
this action as
\begin{multline}
  \label{eq:22}
  I_M=\frac{1}{4\pi}\int d^4x \,\Big[-\cE^i\d_0 A_i+\sqrt g \d^i C
  \d_0 Z_i -A_0\d_i\cE^i+Z_0\d_i\cB^i-\\-\frac{N}{2\sqrt
    g}(\cE^{i}\cE_i+\cB^i\cB_i) +\epsilon_{ijk}N^i\cE^{j}\cB^{k}\Big],
\end{multline}
where $\cE^i=\epsilon^{ijk}\d_j Z_k+\sqrt g \d^i Y$ and
$\cB^i=\epsilon^{ijk}\d_j A_k+\sqrt g\d^i C$.
We assume here and below that every $3$-vector admits a unique
orthogonal, spatially covariant decomposition (see
e.g.~\cite{deser:1967aa}) $X^i=X^{Ti}+X^{Li}$, where
\begin{eqnarray}
  \label{eq:75}
  X^{Li}=\d^i M,\quad X^{Ti}=\frac{1}{\sqrt
  g}\epsilon^{ijk}\d_jL_k,
\end{eqnarray}
for some $M,L_k$. In terms of the inverse of the spatially covariant
Laplacian $\nabla^{-2}$ and the spatially covariant derivative
$\nabla_i$, we have
\begin{eqnarray}
  \label{eq:111}
  M=\nabla^{-2}\nabla_j X^j,\quad  X^{Ti}=X^i-\d^i M.
\end{eqnarray}
A vector is transverse if its divergence vanishes
and longitudinal if its curl vanishes,
\begin{eqnarray}
  \label{eq:76}
\d_i(\sqrt g X^i)=0\Rightarrow X^i=X^{Ti}, \quad
\epsilon^{ijk}\d_j X_k=0\Rightarrow X^{i}=X^{Li}.
\end{eqnarray}
We then have
\begin{equation}
\int d^3x\,\sqrt g X^i g_{ij} Y^j=\int d^3x\,\sqrt g (X^{Li} g_{ij}
Y^{Lj}+X^{Ti} g_{ij}Y^{Tj}).\label{eq:64}\nonumber
\end{equation}

Using such a decomposition for $A^i_a$, $A^a_i=\d_iM^a+A^{aT}_i$,
$M^a=(M_A,M_z)$, the kinetic term becomes
\begin{multline}
  \label{eq:114}
 \int d^4x\, a_A(z)\d_0z^A= \int d^4x\Big(
\big[\frac{\pi^{ij}}{16\pi}+\frac{\sqrt g{D^{ijkl}}}{4\pi}
(Z_k^T\sqrt g\d_l C  -A^T_k\sqrt g\d_l Y)\big]\d_0 g_{ij}-\\-
\frac{\epsilon^{ijk}\d_jZ^T_{k}}{4\pi}\d_0
A^T_i+\frac{\d_i(\sqrt g\d^i Y)}{4\pi}\d_0 M_A
- \frac{\d_i(\sqrt g\d^i C)}{4\pi}\d_0 M_Z\Big),
\end{multline}
where $D_{ijkl}=\frac{1}{2\sqrt g}(g_{ik}g_{jl} +
g_{il}g_{jk}-g_{ij}g_{kl})$ is the DeWitt supermetric. Note that
$D^{ijkl}$ is not the inverse DeWitt supermetric, but the result of
raising all the indices with the metric. Let us define
\begin{eqnarray}
  \label{eq:116}
 \pi^a=(\d_i({\sqrt g\d^i C}),\d_i({\sqrt g\d^i Y}))=(-\pi_Z,\pi_A),
\end{eqnarray}
as new independent variables, so that
$C^a=\nabla^{-2} \frac{\pi^a}{\sqrt g}$.
We also define
\begin{eqnarray}
  \label{eq:123}
  \tilde\pi^{ij}={\pi^{ij}}+4{\sqrt g{D^{ijkl}}}
 \epsilon_{ab} A^{aT}_k\sqrt g \d_l C^b,
\end{eqnarray}
The independent phase space variables are thus
$(g_{ij},A_i^{T},M_A,M_Z,\tilde\pi^{ij},Z_i^T,\pi_A,\pi_Z)$
in terms of which the canonically conjugate pairs are
\begin{eqnarray}
  \label{eq:117}
  (g_{ij},\frac{\tilde \pi^{kl}}{16\pi}),\quad (A_i^T,-
\frac{\epsilon^{ijk}\d_jZ^T_{k}}{4\pi}),\quad
(M_A,\frac{\pi_A}{4\pi}), \quad
(M_Z,\frac{\pi_Z}{4\pi}).
\end{eqnarray}
In particular,
\begin{eqnarray}
\{\cB^{ai}(x),\cB^{bj}(y)\}=4\pi\epsilon^{ijk}
\epsilon^{ab}\d^x_k\delta^3(x,y), \quad
\{M^a(x),\pi^b(y)\}=4\pi\epsilon^{ab}\delta^3(x,y). \label{eq:56}
\end{eqnarray}

\subsection{Gauge structure}
\label{sec:gauge-structure}

Before turning to the equations of motions and their solutions, let us
discuss the gauge structure of the theory. We want to show that the
constraints $\gamma_\alpha$ are first class. Defining
$\epsilon^\alpha\equiv(\xi^\perp,\xi^i,\lambda^a)$ with
$\epsilon^\alpha$ vanishing at the boundary and
$ \Gamma[\epsilon]=\int d^3x\,\gamma_\alpha\epsilon^\alpha$, this
means that
\begin{eqnarray}
  \label{eq:57}
  \{\Gamma[\epsilon_1],\Gamma[\epsilon_2]\}=\Gamma[[\epsilon_1,\epsilon_2]],
\end{eqnarray}
for a suitably defined $[\epsilon_1,\epsilon_2]$. In this case, the
gauge transformations leaving action \eqref{eq:25} invariant are given
by
\begin{eqnarray}
  \label{eq:54}
\delta_\epsilon z^A=\{z^A,\Gamma[\epsilon]\},\quad
\delta u^\alpha=\d_0\epsilon^\alpha+[\epsilon,u]^\alpha.
\end{eqnarray}

In order to compute these brackets it is useful to go to the Darboux
coordinates identified in the previous subsection in terms of which
the Gauss constraints become
\begin{eqnarray}
  \label{eq:119}
\cG_a=\frac{1}{4\pi}\epsilon_{ab}\pi^b.
\end{eqnarray}
Now one should do the change of coordinates in $\cH_\perp,\cH_i$,
i.e., perform the replacement $\pi^{ij}=\tilde\pi^{ij}-4{\sqrt
  g{D^{ijkl}}} \epsilon_{ab} A^{aT}_k\sqrt g \d_l C^b$. Since the
additional terms that are generated in this way are proportional to
$C^a$ and thus vanish on the constraint surface defined by $\cG_a$
they can safely be discarded in the source-free situation. In the
following, we will drop the tilde on $\pi^{ij}$.

In particular, we have
\begin{eqnarray}
\delta_\epsilon \pi^a&=&0,\label{eq:43c}\\
  \label{eq:43}
\delta_\epsilon A^a_i&=&\d_i\lambda^a_T-\frac{g_{ij}}{\sqrt
  g}\epsilon^{ab}
\cB^{j}_b\xi^\perp-\epsilon_{ijk}\xi^j\cB^{ak},\label{eq:supl}
\\
  \delta_\epsilon g_{ij}&=&
 \nabla_i \xi_{j}+\nabla_j\xi_{i}+2D_{ijkl}\pi^{kl}\xi^\perp,\label{eq:43a}
\end{eqnarray}
where
\begin{eqnarray}
  \lambda_T^a=\lambda^a-\epsilon^{ab}\frac{1}{\sqrt
    g}\nabla^{-2}\d_i(\cB^i_b\xi^\perp)+ \frac{1}{\sqrt
    g}\nabla^{-2}\d_i(\frac{\epsilon^{ijk}}{\sqrt g}\cB_j^a\xi_k).
\end{eqnarray}
Equation \eqref{eq:43c} implies that $\pi^a$ are constants of motion,
which is consistent with the longitudinal part of the
source-free Maxwell equations. From \eqref{eq:supl} we also find
\begin{eqnarray}
  \delta_\epsilon\cB^{ia}=-\epsilon^{ijk}\d_j(\frac{1}{\sqrt g}
  \epsilon^{ab}\cB_{bk}\xi^\perp)-\d_j(\cB^{aj}\xi^i)+
\d_j(\cB^{ai}\xi^j).
\label{eq:43b}
\end{eqnarray}
Infinitesimal diffeomorphisms along $\eta^\mu$ are recovered by using
$\xi^\perp=N\eta^0$, $\xi_i=g_{i\mu}\eta^\mu$. Indeed, with this
choice of parameters, $\cL_\eta g_{\mu\nu}\approx
\delta_{\xi}g_{\mu\nu}$. This can be seen for instance on
\eqref{eq:43a} by using the (auxiliary) equations of motion for
$\pi^{ij}$ together with the definitions of lapse $N$ and shift $N^i$
in terms of the $4$-metric $g_{\mu\nu}$. In other words,
diffeomorphism invariance in the Hamiltonian framework is implemented
through the gauge transformations generated by $H[\xi]=\int
d^3x\,(\cH_\perp\xi^\perp+\cH_i\xi^i)$. In particular, using as gauge
parameters the Lagrange multipliers, $\epsilon^\alpha=u^\alpha$,
amounts to performing an infinitesimal time-translation on account of
the Hamiltonian equations of motion.

Let us end this discussion by determining $[\epsilon_1,\epsilon_2]$.
The constraints $\cG_a$, have vanishing Poisson brackets among
themselves and with all other constraints because the $\cB^{ai}$ do
not depend on $M_A,M_Z$. It follows that
$[\lambda,\epsilon_2]^\alpha=0$ and also, from \eqref{eq:43} and
\eqref{eq:54} that the associated gauge transformations
$\delta_\lambda$ generated by $G[\lambda]=\int d^3x\,\cG_a\lambda^a$
are the double electromagnetic gauge transformations of \eqref{eq:45},
while all other variables are left invariant.

We still have to compute $\{H[\xi],H[\eta]\}$. We notice first of all
that the purely gravitational part satisfies the algebra of surface
deformations \cite{Teitelboim:1972vw,Hojman:1976vp},
\begin{eqnarray}
&& \{H^{ADM}[\xi],H^{ADM}[\eta]\}=H^{ADM}[[\xi,\eta]_{SD}],\label{eq:52f}\\
  && [\xi,\eta]^\perp_{SD}=\xi^i\d_i\eta^\perp-\eta^i\d_i\xi^\perp,
\label{eq:52a}\\
  && [\xi,\eta]^i_{SD}=g^{ij}(\xi^\perp\d_j\eta^\perp-\eta^\perp\d_j\xi^\perp)
  +\xi^j\d_j\eta^i-\eta^j\d_j\xi^i.\label{eq:52b}
\end{eqnarray}
{}From \eqref{eq:43b}
and \eqref{eq:43a}, we find that
\begin{multline}
  \label{eq:49}
  \{H^{ADM}[\xi],H^{mat}[\eta]\}-(\xi\leftrightarrow\eta)+
\{H^{mat}[\xi],H^{mat}[\eta]\}=\\=H^{mat}[[\xi,\eta]_{SD}]
+G[[\xi,\eta]_B],
\end{multline}
where
\begin{eqnarray}
  \label{eq:38}
  [\xi,\eta]^a_B=\cB^{ai}\epsilon_{ijk}\xi^j\eta^k-
\frac{\epsilon^{ac}\cB_{ci}}{\sqrt g}
  (\xi^\perp\eta^i -\eta^\perp\xi^i).
\end{eqnarray}
Combining with \eqref{eq:52f}, we finally get
\begin{eqnarray}
  \label{eq:28}
\{H[\xi],H[\eta]\}=H[[\xi,\eta]_{SD}]+G[[\xi,\eta]_B].
\end{eqnarray}

According to \cite{teitelboim:1980aa}, such a constraint algebra
provides the integrability conditions that guarantee that ``the
evolution of a three geometry can be viewed as the deformation of a
three-dimensional cut in a four-dimensional space-time''.

\subsection{Derivation of the Poisson algebra of Poincar\'e generators
  in flat spacetime}
\label{sec:append-poiss-algebra}

In this subsection, we derive the Poisson algebra of the Poincar\'e
generators in flat spacetime as given in \eqref{eq:xx} by restricting
the results of the previous subsection to flat spacetime.

We thus assume in this subsection that
$N=1,N^i=0,g_{ij}=\delta_{ij}$. Greek indices take values from $0$ to
$3$ with $\mu=(\perp,i)$. Indices are lowered and raised with
$\eta_{\mu\nu}={\rm diag} (-1,1,1,1)$ and its inverse. Let
$\tilde\omega_{\mu\nu}=-\tilde\omega_{\nu\mu}$. In this case, the Lie
algebra of vector fields $\xi(\tilde \omega,\tilde a)=(-{\tilde
  \omega^{\mu}}_{\phantom{\mu}i}x^i+\tilde a^{\mu})\ddl{}{x^\mu}$ with
bracket the surface-deformation bracket \eqref{eq:52a} and
\eqref{eq:52b} forms a representation of the Poincar\'e algebra
\cite{Regge:1974zd},
\begin{eqnarray}
  \label{eq:99}
  [\xi(\tilde\omega_1,\tilde a_1),\xi(\tilde \omega_2,\tilde a_2)]_{SD}=
  \xi([\tilde
  \omega_1,\tilde \omega_2],\tilde \omega_1\tilde 
a_2-\tilde \omega_2\tilde a_1).
\end{eqnarray}
It then follows from \eqref{eq:28} that this is also the case for the
canonical generators $\cH[\xi(\tilde \omega,\tilde a)]$ equipped with
the Poisson bracket, when one considers the restriction to the
constraint surface defined by $\cG_a=0$.

Comparing with Sec.~\bref{sec:poinc-gener-their}, we find that
\begin{eqnarray}
  \label{eq:100}
&&  \half\omega_{\mu\nu}J^{\mu\nu}-a_\mu P^\mu=\cH[\xi(\tilde \omega,\tilde
  a)]\nonumber\\
&& \iff \tilde\omega_{\mu\nu}=4\pi\omega_{\mu\nu},\quad \tilde a_0=4\pi
a_0, \quad \tilde a_i=4\pi (a_i+\omega_{0i}x^0).
\end{eqnarray}
This concludes the proof that the generators defined in \eqref{eq:92}
form a representation of the Poincar\'e algebra and the algebra
in \eqref{eq:102} is a direct consequence of \eqref{eq:28}.

\subsection{Equations of motion with sources and comparison to covariant
  formalism}
\label{sec:equat-moti-comp}

The standard Einstein-Maxwell equations, now in the presence of
external, magnetic and electric conserved current densities
$j^{a\mu}$, $\d_\mu j^{a\mu}=0$ given by \eqref{eq:current} with
associated string terms defined in \eqref{sc}, derive
from extremizing the action $I_{geom}+I^\prime_M$, where
\begin{eqnarray}
  \label{eq:59}
  I_{geom}[g_{\mu\nu}]=\frac{1}{16\pi}\int d^4x\,
  \sqrt{-{}^{(4)}g}R,
\end{eqnarray}
and
\begin{multline}
I^\prime_M[g_{\mu\nu},\cF^{\mu\nu},a_\mu,y^\mu]=\frac{1}{4\pi}\int d^4x
  \Big[-\half (\d_\mu a_\nu-\d_\nu
    a_\mu+{}^*G_{\mu\nu})\cF^{\mu\nu}+\\+\frac{1}{4}
    \frac{1}{\sqrt{-{}^{(4)}g}}\cF^{\mu\nu}\cF_{\mu\nu}+
    a_\mu j^\mu\Big]. \label{eq:59a}
\end{multline}
To make connection with the ADM formalism, we have followed
\cite{Arnowitt:1962aa} and introduced the auxiliary tensor densities
$\cF^{\mu\nu}$. On the one hand, one can solve the equations of motion
for $\cF^{\mu\nu}$ algebraically,
$\cF^{\mu\nu}=\sqrt{-{}^{(4)}g}g^{\mu\alpha}g^{\nu\beta}(\d_\mu
a_\nu-\d_\nu a_\mu+{}^*G_{\mu\nu})$. When substituted into the action,
one recovers the Einstein-Maxwell theory with Dirac strings. On the
other hand, one can introduce $\cF^{0i}=\cE^i_{ADM}$, eliminate the
auxiliary $\cF^{ij}$ and use the decomposition of the $4$-metric into
the $3$-metric, lapse and shift to find the standard Hamiltonian form
\begin{multline}
  \label{eq:60}
  I^\prime_M[\cE^i_{ADM},a_\mu,g_{ij},N,N^i,y^\mu]=\frac{1}{4\pi}\int d^4x \,
\Big[-\cE^i_{ADM}(\d_0 a_i+\alpha_i)-a_0\d_i\cE^i_{ADM} -\\-\frac{N}{2\sqrt
    g}(\cE^{i}_{ADM}\cE^{ADM}_i+\cB^i_{ADM}\cB^{ADM}_i)
  +\epsilon_{ijk}N^i\cE^{j}_{ADM}\cB^{k}_{ADM}+a_\mu
  j^{\mu} \Big],
\end{multline}
where $\cB^i_{ADM}=\epsilon^{ijk}\d_ja_k+\beta^k$. The
equations of motion for $\cE^i_{ADM}$ read
\begin{eqnarray}
  \label{eq:61}
  \cE^i_{ADM}=\frac{\sqrt g}{N} g^{ij}(\d_j a_0-\d_0
  a_j-\alpha_j-\epsilon_{jkl} N^k \cB^l_{ADM}).
\end{eqnarray}
They determine $\cE^i_{ADM}$ in terms of the other variables and the
sources. Similarily, in the gravitational sector, the equations of
motion following from varying $\pi^{ij}$ are auxiliary in the sense
that they can be solved algebraically for $\pi^{ij}$ in terms of the
other variables. After this has been done, the constraints and the
equations of motion following from variation of $g_{ij}$ are
equivalent to the covariant Einstein-Maxwell equations with Dirac
strings. Hence, every solution $g_{\mu\nu},a_\mu$ to the covariant
equations of motions is a solution to the ADM equations of motion with
electric and magnetic fields $\cE^i_{ADM},\cB^i_{ADM}$ and momenta
$\pi^{ij}$ determined in terms of $g_{\mu\nu},a_\mu$.  Conversely,
every solution of the ADM equations of motion gives a solution to the
covariant equations of motion.

Alternatively, one can multiply \eqref{eq:61} by $N$. The longitudinal
part of this equation is solved uniquely for $a_0$, while the
transverse part gives, after using
$\epsilon^{ijk}\d_j\alpha_k-\d_0\beta^i=k^i$,
\begin{eqnarray}
  \label{eq:108}
  \d_0\cB^i_{ADM}=-\epsilon^{ijk}\d_j(\frac{N}{\sqrt
    g}\cE_k^{ADM})-\d_j(N^i\cB^j_{ADM}) +\d_j(N^j\cB^i_{ADM})-k^i.
\end{eqnarray}
Finally, Maxwell's equations for $a_i$ are
\begin{eqnarray}
  \label{eq:106}
  \d_0\cE^{i}_{ADM}=\epsilon^{ijk}\d_j(\frac{N}{\sqrt g}
  \cB^{ADM}_k)-\d_j(N^i\cE^{j}_{ADM})+
\d_j(N^j\cE^{i}_{ADM})- j^{i}.
\end{eqnarray}

As a side remark, note that the constraint algebra in the absence of
sources and strings in the standard ADM approach to Einstein-Maxwell
theory can be directly rederived from our result \eqref{eq:28} and is
given by
\begin{eqnarray}
  \label{eq:70}
  \{H[\xi],H[\eta]\}=H[[\xi,\eta]_{SD}]+G_{ADM}[[\xi,\eta]_{ADM}],
\end{eqnarray}
where
\begin{eqnarray}
  \label{eq:69}
  G_{ADM}[\lambda]&=&\int d^3x\,\d_i\cE^i_{ADM} \lambda,\cr
  [\xi,\eta]_{ADM}&=&\cB^{i}_{ADM}\epsilon_{ijk}\xi^j\eta^k-
\frac{\cE^{ADM}_{i}}{\sqrt g}
  (\xi^\perp\eta^i -\eta^\perp\xi^i).
\end{eqnarray}
Indeed, the algebra rests only on the constraints and the
transformation properties \eqref{eq:43a}, \eqref{eq:56}. Provided that
$\cE^i_{ADM}=\cE^i$ and $\cB^i_{ADM}=\cB^i$, which we always assume in
the following, these are the same in both descriptions, except that
the constraint $\cG_2=-\d_i\cB^i_{ADM}$ is absent because
$\d_i\cB^i_{ADM}$ vanishes identically in the absence of sources and
strings.

In the presence of sources, \eqref{d} generalizes readily to curved
space, where a manifestly duality invariant action principle is
defined by $I_{ADM}+\bar I_M$, with
\begin{eqnarray}
  \label{eq:109}
  \bar I_M[a^a_i,g_{ij},N,N^i,y^\mu]=
\frac{1}{8\pi}\int
  d^4x\,\Big[( b^{ai}
\epsilon_{ab}(\d_0 a^b_i+\alpha^b_i)
-\\-\frac{N}{\sqrt g} b^{i}_a b_i^a-\epsilon_{ab}
\epsilon_{ijk}N^i b^{aj} b^{bk}+\epsilon_{ab}a_i^aj^{bi}\Big],
\end{eqnarray}
and $b^a_i=\epsilon^{ijk}\d_ja^a_k+\beta^{ai}$. Indeed, equivalence
of the associated equations of motion to the ADM/covariant ones is
obvious when $b^{1i}=\cB^i_{ADM},b^{2i}=\cE^i_{ADM}$ since the
gravitational equations of motion are unaffected while those for
$a^a_i$ read
\begin{eqnarray}
  \label{eq:110}
   \d_0b^{ai}=-\epsilon^{ijk}\d_j(\frac{N}{\sqrt g}
  \epsilon^{ab}b_{bk})-\d_j(N^ib^{aj})+
\d_j(N^jb^{ai})- j^{ai},
\end{eqnarray}
and coincide with the relevant Eqs.~\eqref{eq:108}-\eqref{eq:106}.

With the longitudinal electric and magnetic fields produced
by the potentials $C^a$, the appropriate action principle is
$I_{ADM}+I_M+I_J$, where $I_{M}$ is defined in \eqref{eq:17} and
\begin{multline}
  \label{eq:85}
  I_J[A_\mu^a,C^a,y^\mu]= \frac{1}{4\pi}\int
  d^4x\,\epsilon_{ab}\big( A_\mu^aj^{b\mu}+ \sqrt g\d^i
  C^a\alpha^b_i-\\-\half \beta^{ai}\alpha^b_i +\half
  \beta^{aTi}\d_0\gamma^{b}_i\big).
\end{multline}
Here $\gamma^{a}_i$ is the potential for the transverse part of
$\beta^{ai}$, $\beta^{aTi}=\epsilon^{ijk}\d_j\gamma^a_k$.

In this case the equations of motion for $A_i^a$ are given by
\begin{eqnarray}
  \label{eq:27}
 \d_0\cB^{ai}=-\epsilon^{ijk}\d_j(\frac{N}{\sqrt g}
  \epsilon^{ab}\cB_{bk})-\d_j(N^i\cB^{aj})+
\d_j(N^j\cB^{ai})- j^{ai}.
\end{eqnarray}
They are the correct matter field equations provided that
$\cB^{ai}=b^{ai}$ are the magnetic and electric fields. This implies
on the one hand
$\epsilon^{ijk}\d_jA_k^a=\epsilon^{ijk}\d_ja_k^a+\beta^{aTi}$, and in
turn $A^{a}_k=a^{a}_k+\gamma^a_k$, up to an irrelevant longitudinal
part, and $\sqrt g\d^i C^a=\beta^{aLi}$ on the other hand. Again, the
equations of motion for $A_0^a,C^a$,
\begin{eqnarray}
  \label{eq:50}
&&\partial_i(\sqrt g\partial^i C^a)=j^{a0},\\
&& \partial_i\Big[N \cB^i_a-\epsilon_{ab}
  \sqrt g g^{il}\big(\d_0 A^b_l-\d_l A^b_0+\alpha_l^b+\epsilon_{ljk}
  N^j\cB^{bk}\big)\Big]=0,
\end{eqnarray}
are auxiliary because they can be used to solve these fields in terms
of the others. This can be done in the action principle and gives back
\eqref{eq:109}.

In conclusion, if the lapse $N$ is nonvanishing and the covariant
decomposition of spatial vectors into longitudinal and transverse
components is unique then there is a one-to-one and onto
correspondence between solutions of the covariant/ADM equations of
motion and solutions to the equations of motion deriving from
$I_{ADM}+I_M+I_J$.

In the case of a single dyon, one can again drop all string terms in
$I_J$, which then simplifies to
\begin{eqnarray}
I_J[A_\mu^a;j^{a\mu}]= \frac{1}{4\pi}\int
  d^4x\,\epsilon_{ab} A_\mu^aj^{b\mu}\label{eq:112}.
\end{eqnarray}
All relevant matter equations of motion are correct in this case, but
one has to face the fact that the metric dependence in the
longitudinal part of $\cB^{ai}$ implies an additional term in the
equations of motion associated with variations of $g_{kl}$,
\begin{eqnarray}
  \label{eq:66}
  \vddl{(I_{ADM}+I_M)}{g_{kl}}-\vddl{(I_{ADM}+\bar I_M)}{g_{kl}}  =
  \frac{\sqrt g}{4\pi}D^{ijkl}\d_j
  C^a X_{ai},\\
  X^i_{a}=\frac{N}{\sqrt g}\cB^i_{a}-
\epsilon_{ab}{g^{il}}\big(\d_0 A^b_l-\d_l A_0^b+\epsilon_{lkm}N^k\cB^{bm}\big).
\end{eqnarray}
Again, one can use a duality rotation to make the magnetic charge
vanish in which case the equations of motion imply $C^1=0$. We thus
only need to consider $X^i_2$. But in the purely electric
case $\alpha_i=0$ and $A_\mu=a_\mu$ so that $X^i_2$ vanishes on
account of the matter equation of motion \eqref{eq:61}.

\subsection{String-singularity free dyonic black holes }
\label{sec:string-sing-free}

Consider now the case of a dyonic Reissner-Nordstr{\o}m solution with
charge $Q^a$.  The dyon defined by \eqref{eq:20}-\eqref{eq:20a} is a
solution outside of the location of the dyon (at $r=0$ in our
coordinate system) with a Dirac-string singularity to the equations
deriving from $I_{geom}+I^\prime_M$ given in \eqref{eq:59} and
\eqref{eq:59a}. It is thus also a solution to the equations of motion
derived from $I_{ADM}+ I^\prime_M$ for which
\begin{eqnarray}
  \label{eq:62}
  \cE^i_{ADM}=\delta^i_rQ\sin\theta ,\quad \cB^i_{ADM}=\delta^i_r P\sin\theta.
\end{eqnarray}

In the simplified duality invariant formulation defined by
$I_{ADM}+I_M+I_J$, with $I_J$ given in \eqref{eq:112}, we have to
determine the vector and scalar potentials giving rise to
$\cB^{ai}=\delta^i_rQ^a\sin\theta$, where $Q^1=P, Q^2=Q$ with the
metric given by \eqref{eq:20}. This is easily seen to be the case for
$A^{ai}=0$ and
\begin{eqnarray}
  \label{eq:77}
   C^a
&=&
-Q^a \int^\infty_{r} \frac{ dr^\prime}{{r^\prime}^2 N(r^\prime)}
\nonumber\\ &=&
 \frac{ Q^a}{\sqrt{Q^b Q_b}}
  \ln{\frac{r(M-\sqrt{Q^fQ_f})}{Mr-Q^cQ_c-\sqrt{Q^dQ_d(r^2-2Mr+
        Q^eQ_e)}}}\nonumber\\
  &=&
-\frac{Q^a}{r}+O(r^{-2}).\label{eq:63a}
\end{eqnarray}
In the gauge where the scalar potentials vanish at infinity, it is
then straightforward to see that all matter equations of motions are
solved by
\begin{eqnarray}
  A^a_0&=&-\frac{\epsilon^{ab}Q_b}{r},\label{eq:63b}
\end{eqnarray}
and one can directly check that in this case $X^i_a=0$. In conclusion,
in the new formulation, the Reissner-Nordstr{\o}m dyon is described by
the metric \eqref{eq:20} and the potentials \eqref{eq:77},
\eqref{eq:63b}. The string singularity of the standard approach has
thus been resolved in the new formulation.

In the gauge where the scalar potentials vanish at infinity, let us
define
\begin{eqnarray}
\phi=-A_0,\quad \psi=Z_0\label{eq:1}.
\end{eqnarray}
with $\phi_H,\psi_H$ denoting these quantities evaluated at the horizon, in agreement with \eqref{eq:35}.
For the resolved Reissner-Nordstr{\o}m dyon, this gives
\begin{eqnarray}
  \label{eq:78}
  \phi=\frac{Q}{r},\quad \psi=\frac{P}{r}.
\end{eqnarray}
In the Euclidean methods discussed below, it is useful to choose a
gauge where the scalar potentials vanish on the horizon. In this case,
\begin{eqnarray}
  A^a_0=-\epsilon^{ab}Q_b(\frac{1}{r}-\frac{1}{r_+}),\\
A_0=-\frac{Q}{r}+\phi_H,\quad Z_0=\frac{P}{r}-\psi_H.
\label{eq:63}
\end{eqnarray}

\subsection{Surface charges}
\label{sec:surf-charg-first}

The Regge-Teitelboim analysis \cite{Regge:1974zd} allows one to derive
the correct variational principle in the presence of nonvanishing
surface charges at infinity.  Consider an arbitrary gauge
transformation generator $\Gamma[\epsilon]= \int d^3 x
\gamma_\alpha\epsilon^\alpha$ with $\epsilon^\alpha$ not necessarily
vanishing at the boundary (a condition we demanded in Section 4.2
above). The variation of this generator under a change of phase space
variables may be written
 \begin{eqnarray}
  \label{eq:29}
\delta_z \Gamma[\epsilon] =
\int d^3x\, \delta_z(\gamma_\alpha \epsilon^\alpha)=\int d^3x \left(\delta
z^A\vddl{(\gamma_\alpha\epsilon^\alpha)}{
    z^A}-\d_i k^{i}_{\epsilon}\right),
\end{eqnarray}
where $\delta/\delta z^A$ is the Euler-Lagrange derivative. The second
piece is a boundary term and arises from integration by parts. The
expression $k^{i}_\varepsilon[z^A,\delta z^A]$ depends on the phase
space variables and linearly on their variations and the gauge
parameters.

For the simplest application, consider phase space variables $z^A_s$
that satisfy the constraints and variations $\delta
z^A_s$ obeying the linearized constraints. In this case, the left
hand side of \eqref{eq:29} vanishes. Suppose then that the associated
solution $z^a_s,u^\alpha_s$ to the evolution equations is time
independent, $\d_0z^A_s=0$. In particular, this means that the
associated vector field $ \eta^\mu=\delta^\mu_0$ is the timelike
Killing vector field of the metric $g_{\mu\nu}$. In this case, the
evolution equations following from \eqref{eq:25},
\begin{eqnarray}
  \label{eq:41}
  \vddl{a_B}{z^A}\d_0z^A-\d_0 a_A=\vddl{(\gamma_\alpha u^\alpha)}{z^A},
\end{eqnarray}
imply that the first term on the right hand side of \eqref{eq:29}
vanishes as well. We thus find
\begin{eqnarray}
  \label{eq:19}
  \d_i k^{i}_{u_s}[z^A_s,\delta z^A_s]=0.
\end{eqnarray}
Using Stokes' theorem, it follows that the integral over a sphere at
radius $r$ and fixed time $t$ does not depend on the radius $r$,
\begin{eqnarray}
  \label{eq:3}
  \oint_{S_{r_1}} d^{3} x_i\ k^{i}_{u_s}[z^A_s,\delta z^A_s]=
\oint_{S_{r_2}} d^{3} x_i\ k^{i}_{u_s}[z^A_s,\delta z^A_s].
\end{eqnarray}
Here $d^{3}x_i=\half\epsilon_{ijk}dx^j\wedge dx^k$.  The explicit
expression for $k^{i}_\epsilon[z^A;\delta z^A]$ can be easily worked
out by integrations by parts. It is defined up to the divergence of an
arbitrary superpotential, $\partial_j t^{[ij]}$, which does not play
any r\^ole for our purpose. It splits into a standard purely
gravitational part and a matter part,
\begin{eqnarray}
k^{i}_{\epsilon}[z^A;\delta z^A]=k^{grav,
  i}_{\epsilon}[g_{ij},\pi^{ij};\delta g_{ij},\delta\pi^{ij}]
+k^{mat, i}_{\epsilon}[z^A;\delta z^A].\label{eq:29a}
\end{eqnarray}
The former has been derived in \cite{Regge:1974zd} and reads
\begin{eqnarray}\begin{array}{l}
  \label{eq:21b}
  k^{grav, i}_{\epsilon}=\frac{1}{16\pi}\Big[G^{ljki} (\xi^\perp\nabla_k\delta
  g_{lj}-\d_k\xi^\perp\delta
  g_{lj})+2\xi_k\delta\pi^{ki}+(2\xi^k\pi^{ji}-\xi^i\pi^{jk}) \delta
  g_{jk}\Big],\\
G^{ljki}=\half\sqrt g(g^{lk}g^{ji}+g^{il}g^{jk}-2g^{lj}g^{ki}),
\end{array}
\end{eqnarray}
where $G^{ijkl}$ is the inverse of the DeWitt supermetric,
$D_{ijkl}G^{klmn}=\half (\delta^m_i\delta^n_j+\delta^m_j\delta^n_i)$.
The matter part now involves, besides the electric contributions, the
sought for magnetic ones:
\begin{multline}
  k^{mat, i}_{\epsilon} =
  \frac{1}{4\pi}\Big(\frac{\xi^\perp}{\sqrt{g}}\epsilon^{ijk}\cB^a_j\delta
  A_{ak}-\xi^\perp \cB^{ai} \delta
  C_a+\epsilon_{ab}(\xi^k\cB^{ai}-\xi^i \cB^{ak})\delta A^b_k-\\ -
  \epsilon_{ab}\sqrt g g^{il}\epsilon_{ljk}\xi^j\cB^{ak}\delta C^b
  +\epsilon_{ab}(\sqrt g
  \d^i\lambda^a\delta C^b -\lambda^a\delta\cB^{bLi})\Big).\label{eq:52d}
\end{multline}
In the sequel, we are interested in asymptotically flat gravitational
field configurations carrying finite charges associated with
energy momentum. We will not need to consider the more general
boundary conditions guaranteeing finite charges associated with
rotations or boosts. The appropriate fall-off conditions on the
gravitational variables and lapse and shift have been discussed in
detail in \cite{Regge:1974zd},
\begin{eqnarray}
  \label{eq:21}
&& \hspace*{-1.3cm} g_{rr}=1+O(r^{-1}),\ g_{\theta\theta}=r^2+O(r),\
    g_{\phi\phi}=r^2\sin^2\theta + O(r),\\
&& \hspace*{-1.3cm} g_{r\theta}=O(r^0)=g_{r\phi},\ g_{\theta\phi}=O(r),\\
&&\hspace*{-1.3cm}  \pi^{rr}=O(r^{0}),\ \pi^{\theta\theta}=O(r^{-2})=
\pi^{\phi\phi}=\pi^{\theta\phi},\ \pi^{r\theta}=O(r^{-1})=\pi^{r\phi},\\
&&\hspace*{-1.3cm}  N=1+O(r^{-1}),\ N^\phi=O(r^{-2})=
N^\theta,\ N^r=O(r^{-1}).\label{aaaa}
\end{eqnarray}
For the matter variables, we assume
\begin{eqnarray}
  \label{eq:31}
  A_r^a=O(r^{-1}),\ A^a_\theta=O(r^0)=A^a_\phi,\ C^a=O(r^{-1}),\
A^a_0=k^a+O(r^{-1}).
\end{eqnarray}
In particular, these fall-off conditions include the background
solutions $\bar z,\bar u$ described by
\begin{eqnarray}
\bar g_{rr}=1,\ \bar g_{\theta\theta}=r^2,\ \bar
g_{\phi\phi}=r^2\sin^2\theta,\\
\bar N=1,\ \bar N^\phi=0=\bar N^\theta=\bar N^r,\ \bar A_0^a=k^a,
\label{eq:31a}
\end{eqnarray}
and all other variables vanishing. For later use we introduce the
additional notation
\begin{eqnarray}
  \label{eq:71}
  k^1=\phi^c,\quad k^2=-\psi^c.
\end{eqnarray}
In order to allow configurations
satisfying the fall-off conditions to be extrema of the variational
principle, action \eqref{eq:25} needs to be supplemented by the
addition of a suitable surface term at the boundary at infinity, i.e.,
the surface $r,t$ constant with $r\rightarrow \infty$,
\begin{eqnarray}
  \label{eq:44}
  I^T[z,u]=\int d^4x
  [a_A(z)\d_0z^A-u^\alpha\gamma_\alpha]- Q_{u}[z].
\end{eqnarray}
The surface term $Q_{u}[z]$ is determined by the requirement
that, under variations of the fields $z^A$ satisfying the fall-off
conditions, its variation $\delta_z Q_{u}$ should precisely
cancel the spatial boundary term arising when deriving the Hamiltonian
equations of motion, i.e., the term due to the right hand side of
\eqref{eq:29},
\begin{eqnarray}
  \label{eq:53}
\delta_z Q_{u}[z]=\oint_{S^\infty}d^3x_i\,
k^i_{u}[z,\delta z].
\end{eqnarray}
For the purely gravitational part, this problem was solved in
\cite{Regge:1974zd}, the appropriate boundary term being the ADM mass:
\begin{eqnarray}
  \label{eq:13}
\hspace*{-.5cm}\oint_{S^\infty}d^3x_i\,
k^{grav,i}_{u}[z,\delta z]=\oint_{S^\infty}d^3x_i\,k^{grav,i}_{\bar
  u}[\bar z,\delta z] =\delta_z \oint_{S^\infty}d^3x_i\,k^{grav,i}_{
\bar u}[\bar z,z-\bar z],
\end{eqnarray}
so that
\begin{eqnarray}
  \label{eq:15}  Q^{grav}_{u}[g,\pi]=
\oint_{S^\infty}d^3x_i\,k^{grav,i}_{\bar
  u}[\bar z,z-\bar z]=\cM ,\\
\cM=\oint d^3x_i\,\sqrt{\bar g} (\bar g^{lk}\bar g^{ji}-\bar
g^{lj}\bar g^{ki})\bar D_k (g_{lj}-\bar g_{lj}) ,
\end{eqnarray}
where the covariant derivative is taken with respect to the flat
background metric $\bar g_{ij}$.

For the matter part, the boundary conditions imply also that
\begin{eqnarray}
\hspace*{-.5cm}\oint_{S^\infty}d^3x_i\,
k^{mat,i}_{u}[z,\delta z]=\oint_{S^\infty}d^3x_i\,k^{mat,i}_{\bar
  u}[\bar z,\delta z] =\delta_z \oint_{S^\infty}d^3x_i\,k^{mat,i}_{\bar
  u}[\bar z,z-\bar z].
\end{eqnarray}
In particular, the boundary conditions \eqref{eq:21}-\eqref{aaaa} are
such that, when $(\xi^\perp,\xi^i,\lambda^a)$ are replaced by
$(N,N^i,A^a_0)$, the contributions proportional to $N,N^i$ from the
matter part \eqref{eq:52d} vanish. There is thus no correction to the
ADM mass for the adopted boundary conditions. This will not remain
true for more general boundary conditions where the matter part
\eqref{eq:52d} can contribute both to the ADM energy momentum and the
Lorentz generators. For the boundary conditions at hand, only the last
term survives and combines into magnetic and electric charge
$\cQ^a=(\cP,\cQ)$, as expected,
\begin{eqnarray}
  \label{eq:58}
 &&  Q^{mat}_{u}[g,C]=\oint_{S^\infty}d^3x_i\,k^{mat,i}_{\bar
  u}[\bar z,z-\bar z]=-k^a\epsilon_{ab}\cQ^b,\\
&& \cQ^b=\frac{1}{4\pi}\oint_{S^\infty} d^3x_i\, \cB^{biL}.
\end{eqnarray}

In other words, off-shell, the correct Hamiltonian for the boundary
condition under considerations is
\begin{eqnarray}
  \label{eq:67}
  \bundle{H}=\int d^3x\, (\cH_\perp N+\cH_i N^i) +\cM,
\end{eqnarray}
while electric and magnetic charges are given by
\begin{eqnarray}
  \label{eq:79}
  \bundle{Q}=-\frac{1}{\phi^c}\int d^3x\, (\cG_1 A_0)+\cQ,\
\bundle{P}=-\frac{1}{\psi^c}\int d^3x\, (\cG_2Z_0) +\cP.
\end{eqnarray}
These obervables commute in the Poisson bracket,
\begin{eqnarray}
  \label{eq:80}
  \{\bundle{H},\bundle{Q}\}=0=\{\bundle{H},\bundle{P}\}=
\{\bundle{Q},\bundle{P}\},
\end{eqnarray}
and the total action \eqref{eq:44} can be written as
\begin{eqnarray}
  \label{eq:4}
  I^T[z,u]=\int dt\, \Big(\int d^3x\,
  a_A(z)\d_0z^A-(\bundle{H}-\phi^c\bundle{Q}-\psi^c\bundle{P})\Big).
\end{eqnarray}

\subsection{First law}
\label{sec:first-law}

For the resolved Reissner-Nordstr{\o}m dyon $z,u$ given by
\eqref{eq:20}, \eqref{eq:63a}, \eqref{eq:63}, the first law of
thermodynamics can now be derived as a consequence of using identity
\eqref{eq:3} between infinity $r_1\to\infty$ and the outer
horizon $r_2=r_+$,
\begin{eqnarray}
  \label{eq:51}
\oint_{S^\infty}d^3x_i\,k^{i}_{u}[z,\delta z]=
\oint_{S_{r_+}}d^3x_i\,k^{i}_{u}[z,\delta z],
\end{eqnarray}
where $\delta_z$ describes a variation around the dyon satisfying
the constraints. Indeed, in this case, $k^\phi=0=k^\theta=k^r$, while
$k^a=\frac{\epsilon^{ab}Q_b}{r_+}$. In other words $k^1=\phi_H$ is the
electric potential on the horizon, while $k^2=-\psi_H$ is minus the
magnetic potential on the horizon. Now, the results of the previous
subsection imply that we get at infinity,
\begin{eqnarray}
  \label{eq:65}
  \oint_{S^\infty}d^3x_i\,k^{i}_{u}[z,\delta z]=
\oint_{S^\infty}d^3x_i\,k^{i}_{\bar u}[\bar
z,z-\bar z]=\delta_z\cM-\phi_H\delta_z\cQ-\psi_H\delta_z
\cP.
\end{eqnarray}

For the matter part, we have
\begin{eqnarray}
  \label{eq:53a}
\oint_{S_r} d^{n-1} x_i\ k^{mat, i}_{u}
 = -\frac{1}{4\pi}\int^{\pi}_0d\theta\,
\int^{2\pi}_0 d\phi\
\epsilon_{ab} A_0^a \delta\cB^{bLi},
\end{eqnarray}
which vanishes on the horizon $r=r_+$ where $A_0^a$ vanishes. Note
that in the gauge where $A_0^a$ vanishes at infinity, the matter part
gives no contribution at infinity, but $\phi_H\delta_z\cQ+\psi_H\delta_z
\cP$ at the horizon, as it should.

Finally at the horizon, the purely gravitational part gives
\begin{eqnarray}
  \label{eq:68}
  \oint_{S_{r_+}} d^{n-1}x_i\, k^{grav, i}_u=\frac{\kappa}{8\pi}\delta_z\cA.
\end{eqnarray}
This can be shown for instance by using the fact that $k^{grav,i}_u$
is the time-space component of a conserved superpotential
$k^{[\mu\nu]}_{\d/\d t}$ that can be proved to coincide, for
variations satisfying the linearized field equations and up to an
irrelevant term of the form $\d_\sigma t_{\d/\d t}^{[\sigma\mu\nu]}$, with
the conserved superpotential considered in \cite{Iyer:1994ys}. In turn
the latter has been shown to contribute
$\frac{\kappa}{8\pi}\delta_z\cA$ at the horizon.

This concludes the geometric discussion of the first law
\begin{eqnarray}
  \label{eq:72}
  \delta_z\cM=\frac{\kappa}{8\pi}\delta_z\cA+
\phi_H\delta_z\cQ+\psi_H\delta_z\cP,
\end{eqnarray}
for variations satisfying the linearized equations of motion
around the Reissner-Nordstr{\o}m dyon.

\subsection{Euclidean approach}
\label{sec:eucl-appr-black}

Our setup also allows us to complement the work of
\cite{Hawking:1995ap},\cite{Deser:1997xu} by evaluating the partition
function in the grand canonical ensemble, along the lines of
\cite{Gibbons:1976ue}.

For the three commuting observables $\bundle{\hat H},\bundle{\hat
  Q},\bundle{\hat P}$, we thus would like to compute
\begin{eqnarray}
  \label{eq:81}
  Z[\beta,\phi^c,\psi^c]=\mathrm{Tr}\, e^{-\beta (\bundle{\hat
      H}-\phi^c\bundle{\hat Q}-\psi^c\bundle{\hat P})} = e^{\Psi_G},
\end{eqnarray}
where $\Psi_G(\beta,-\beta\phi^c,-\beta\psi^c)$ is the Massieu potential
for the grand canonical ensemble
(see e.g.~\cite{0264-9381-7-8-020} in the present context),
\begin{eqnarray}
&& \Psi_G(\beta,-\beta\phi,-\beta\psi)=S(\langle \hat H\rangle, \langle
\hat Q\rangle,\langle \hat P\rangle)-\beta\langle \hat
H\rangle+\beta\phi \langle
\hat Q\rangle+\beta\psi  \langle \hat P\rangle,\\
&& d\Psi_G=-\langle \hat H\rangle d\beta+\langle
\hat Q\rangle d(\beta\phi)+ \langle
\hat P\rangle d(\beta\psi).
  \label{eq:84}
\end{eqnarray}
The path integral representation for this partition
function is
\begin{equation}\label{pf}
 Z[\beta,\phi,\psi] = \int {\cal D} \Phi e^{I^T_e} \ ,
\end{equation}
where $\Phi$ represents all the fields $(z^A,u^\alpha)$ together
with appropriate ghost fields $\bar C^\alpha,C^\alpha$
\cite{Faddeev:1969su} (see e.g.~\cite{Henneaux:1992ig} for a review).
The appropriate action is,
\begin{eqnarray}
  \label{eq:86}
  I^T_e=\int_0^\beta d\tau\Big(i\int d^3x\,
  a_A(z)\d_0z^A -(\bundle{H}-\phi^c\bundle{Q}-\psi^c\bundle{P})\Big)
  +\mathrm{ghost\ terms}.
\end{eqnarray}
The path integral is taken over all periodic paths in $ \tau$ with
periodicity $\beta$ and $N\to 1$, $A_0\to\phi^c$, $Z_0\to -\psi^c$ for
$r\to \infty$.

We now notice that the transformation defined by
\begin{eqnarray}
  \label{eq:83}
\pi^{ij}\to -i\pi^{ij},\ A_i\to -i A_i,\ Z_i^L\to -iZ_i^L,\ N^i\to
-iN^i,
\end{eqnarray}
with all other variables unchanged maps the action $I^T_e$ to a real
action when all (transformed) variables are real. The latter action
differs from the Lorentzian action \eqref{eq:4} by the fact that the
terms involving $\pi^{ij}$ and $\cB^{iT}$ in $N\cH_\perp$ and
$N^i\cH_i$ have the opposite signs. For the purely gravitational part,
this is as it should be in order that the path integral corresponds to
one over Euclidean metrics after integration over the momenta
$\pi^{ij}$.

The leading contribution to the path integral is given by the value of
$e^{I_e^{T}}$ evaluated at the classical solutions satisfying the
specified boundary conditions, that is, (i) the fall-off conditions
\eqref{eq:21}-\eqref{eq:31}, (ii) fixed values of the potentials
$(\phi^c,\psi^c)$, and (iii) a fixed inverse temperature $\beta$.  The
Reissner-Nordstr{\o}m dyon (RND) described by the lapse and the
spatial metric given in \eqref{eq:20} and the matter fields
\eqref{eq:63a}, \eqref{eq:63} is such a solution if $\phi^c=\phi_H$
and $\psi^c=\psi_H$ with $\phi_H,\psi_H$ defined in \eqref{eq:35}
since all variables affected by the above transformation vanish in
this case. Furthermore, this solution is time independent and
satisfies the (modified) constraints so that $I_e^T$ reduces to
surface integrals. For the matter part, we find directly that
$I^{mat}_e(RND)=\beta\phi_HQ+\beta\psi_HP$. For the gravitational
part, it has been shown for instance in \cite{Banados:1993qp} in the
current Hamiltonian context that $I^{grav}_e(RND)=-\beta
M+\frac{1}{4}\cA$, with $\cA$ given in \eqref{eq:35a}.

Assuming then that the dyon is the only extremum, it follows that to
leading order,
\begin{eqnarray}
  \label{eq:82}
\Psi_G=-\beta M+\frac{1}{4}\cA+\beta\phi_HQ+\beta\psi_HP,
\end{eqnarray}
which is the expected result.

\section{Conclusion}
\label{sec:conclusion}

In this paper we have generalized the manifestly duality invariant
double potential formalism \cite{Deser:1976iy}, \cite{Schwarz:1993vs}
to include potentials for the longitudinal electric and magnetic
fields thus turning the scalar potentials into non spurious Lagrange
multipliers. By introducing additional pure gauge degrees of freedom
on the classical level, which corresponds to an additional quartet
\cite{Kugo:1979gm} on the quantum level, we have turned a topological
conservation law, the magnetic charge, into a dynamical one.

We have shown on the example of the Reissner-Nordstr{\o}m dyon that
the formalism is tailor-made for a treatment of black hole dyons by
standard action based methods and allows one to compute in the grand
canonical ensemble. How to explicitly resolve the string-singularity
of the Kerr-Newman dyon and derive its thermodynamics will be
discussed elsewhere.

In our approach Dirac strings are only needed for the coupling to
dynamical dyons and the derivation of the Lorentz force law. It would
be interesting to understand whether there are applications of the
formalism in the non-Abelian case or extensions to gravitational
magnetic charge.

{\bf Note: } After the present work has been accepted for publication,
references \cite{Singleton:1995xu,Singleton:1997hf} have been called
to our attention. In these references, a manifestly covariant double
potential formalism is developed. As explained there, the problem is
the occurence of a second ``photon'' that has to be removed in an ad
hoc manner. 

\section*{Acknowledgements}
\label{sec:acknowledgements}

The authors thank G.~Comp\`ere for useful discussions. A.G.~is
grateful to the International Solvay Institutes for hospitality during
various stages of this project. This work is supported in part by a
``P{\^o}le d'Attraction Interuniversitaire'' (Belgium), by
IISN-Belgium, convention 4.4505.86, by the Fund for Scientific
Research-FNRS (Belgium), by Proyectos FONDECYT 1051084, 7070183, and
1051064, by the research Grant No.~26-05/R of Universidad Andr\'es Bello
and by the European Commission programme MRTN-CT-2004-005104, in which
G.B.~is associated with V.U.~Brussel.


\def\cprime{$'$}
\providecommand{\href}[2]{#2}\begingroup\raggedright\endgroup

\end{document}